\documentclass{article}

\usepackage{cite}
\usepackage{graphicx}
\usepackage{subfigure}
\usepackage{url}

\begin{document}

\title{On the Cooperation of Independent Registries}
\date{December 31, 2008}
\author{Matteo~Miraz}

\markboth{Technical Report}{A Distributed Approach for the Federation of Heterogeneous Registries.}

\maketitle

\begin{abstract}
Registries play a key role in service-oriented applications. Originally, they were neutral players between service providers and clients. The UDDI Business Registry (UBR) was meant to foster these concepts and provide a common reference for companies interested in Web services. The more Web services were used, the more companies started create their own “local” registries: more efficient discovery processes, better control over the quality of published information, and also more sophisticated publication policies motivated the creation of private repositories.
The number and heterogeneity of the different registries -- besides the decision to close the UBR— are pushing for new and sophisticated means to make different registries cooperate. This paper proposes DIRE (DIstributed REgistry), a novel approach based on a publish and subscribe (P/S) infrastructure to federate different heterogeneous registries and make them exchange information about published services. The paper discusses the main motivations for the P/S-based infrastructure, proposes an integrated service model, introduces the main components of the framework, and exemplifies them on a simple case study.
\end{abstract}


\section{Introduction}
\label{sec:intro}
Service-oriented architectures exploit \textit{registries} to expose services to possible clients.  Originally, the registry was a \textit{neutral} actor between clients and providers. It was a ``shared'' resource aimed at facilitating their cooperation. This was the original mission of the first version of the UDDI (Universal Description, Discovery, and Integration, \cite{uddi2spec}) specification, which was the first market-supported standard that allowed companies to publish their services and interact with clients~\cite{libroJWS}. To this end, in September 2000, BEA, IBM, and Microsoft started \textit{UBR} (UDDI Business Registry), a public UDDI-based registry, but also a common and neutral reference for all the companies interested in publishing and exploiting Web services. 
As proposed in 
The diffusion of Web services led to the need for ``private'' registries, directly controlled by the different companies, in parallel with the public one. Companies want to be able to control their registries to increase the efficiency of the discovery process, 
but they also want to manage private information ---e.g., exclusive offers to dedicated clients. In this phase, these registries do not substitute the central one, which continues to be a universally-known reference. If a company is not able to serve a request internally, it can always access the central repository to find the services it needs. 

Both the second version of the UDDI specification~\cite{uddi2spec}, and other approaches, like ebXML~\cite{ebXML}, take a more decentralized view and allow for the creation of registries separated from the central system. Moreover, January this year, the companies behind UBR decided to shut it down~\cite{uddishutdown} since the original goal was met---the creation of a common sandbox to foster the diffusion of the service-oriented paradigm. The new advice is to install a dedicated repository for each company. The complete control over published information allows the company to select and filter the information they want to publish, organize it the way they prefer, and thus better tune the discovery process. Usually, companies manage their services, and those provided by their partners, but the lack of a common search space hinders the discovery of new services ---supplied by providers with which the company is not used to cooperate with. Companies interested in new services must a-priori select the companies that might provide them, and then search their proprietary registries, if allowed. Moreover, clients do not often know the services that fit their needs directly, but they would like to query the registry to find those that better fit their expectations. The more accurate the descriptions associated with services are, the more precise the discovery can be.

To overcome the lack of a centralized repositories, and also to supply an extensible model to describe services, this paper proposes \textit{DIRE} (DIstributed REgistry), a novel approach for the seamless cooperation among registries.
\textit{DIRE} fosters the integration of registries based on different standards (e.g., UDDI, ebXML, etc.) by means of dedicated plugs called \textit{delivery managers}. They adopt a unique service model that both extends and abstracts the model used by the single registries, and provides a flexible means for the characterization of services.

\textit{DIRE} proposes a decoupled approach using a global communication system, that allows each delivery manager to share service descriptions. The core of this communication system is based on the publish and subscribe (P/S) middleware \textit{REDS}~\cite{reds}. A unique distributed dispatcher~\cite{DistributedPS} supports the information exchange among the different registries. Even if the dispatcher is logically unique, it provides a physically distributed communication bus to allow registries to publish information about their services and clients, which may be other registries or suitable application interfaces, to register for specific services, or for services with particular characteristics.

\textit{DIRE} allows one to manage two main kinds of cooperation: the \textit{marketplace} and \textit{federation}.
Marketplace allows service providers and potential customers to exchange service information. The former are able to advertise descriptions about their services, while the latter are able to express their service requirements. \textit{DIRE} is able to forward services descriptions to interested parties, matchmaking users' needs against service features. 
The creation of dedicated \textit{federations} allows for the re-distribution of interesting information  within those registries that belong to the federation. 

The main contribution of \textit{DIRE} are:
\begin{itemize}
\item The definition of a \textit{service model}, independent from a particular registry implementation, able to distinguish between descriptions provided by the service provider from those attached by users.
\item The proposal of \textit{Marketplace}, an innovative way to let independent registries cooperate. It ensures to differentiate between shared descriptions and those that the organisation wants to keep private. 
\item The proposal of \textit{Federation}s, able to group together organisations with similar interests.
\item The realisation of a prototype\footnote{The \textit{DeliveryManager} prototype is publicly available at \url{http://code.google.com/p/delivery-manager/}.}, able to keep separated marketplace and federation concepts from the specific communication protocols. 
We have proved the validity of the overall approach on top of this prototype, measuring some performance indexes.
\end{itemize}

The rest of the paper is organised as follow. 
Section~\ref{sec:relatedWork} surveys similar proposals and paves the ground to our approach. 
Section~\ref{sec:model} describes the technology-agnostic service model defined to document managed services and provide delivery managers with a common base.
Section~\ref{sec:caseStudy} presents the running example, inspired from a real-world scenario, which explains how the delivery manager works.
Section~\ref{sec:deliveryManager} presents the delivery manager and digs down into the mechanisms offered to support flexible cooperation among registries.
Section~\ref{sec:performance} is still under measurement, and will be delivered as soon as possible. 
Section~\ref{sec:conc} concludes the paper summarising achieved results.



\section{Related Work}
\label{sec:relatedWork}
Service discovery is a complex problem that, even if it has gained attention from researchers all over the world, still lacks of a clear solution. 
As explained in \cite{garofalakis2004wsd}, the main obstacle affecting web service discovery mechanisms is heterogeneity, that might be technological (e.g. different platforms or different data formats), ontological (domain-specific terms and concepts within services that can differ from one another, especially when developed by different vendors) and pragmatic (different development of domain-specific processes and different conceptions regarding the support of domain-specific tasks).

Web service discovery approaches can broadly be classified \cite{ReazAhmedPHDThesis} as \textit{centralized} and \textit{decentralized}.

Centralized approaches conceive web service discovery architecture as a dedicated entity that maintains the whole directory information and takes care of registering services and answering to queries. Centralized repositories might be designed as a \textit{registry} or as an \textit{index}.
In the first case, a Universal Business Registry (UBR) represents the authoritative, centrally controlled store of information. Moreover an UBR can reference information store elsewhere, for example in a detached repository. The registry owner has to decide who has authority to place information into, or update, the registry or delegate permission to approved provider entities that wish to publish their own service descriptions. Publishing a service description requires an active step by the provider entity: it must explicitly place the information into the registry before that information is available to others. Since a centralized registry might be a bottleneck and a single point of failure, this approach didn't demonstrate to scale well and to be robust as the number of services grows. 

An index, instead, is a reference to information that exists elsewhere. It is not authoritative and does not centrally control the information that it references. Publishing is passive, since are the index owners to collect functional and service description exposed by providers or anyone on the web, without provider entity's specific knowledge. Since anyone can create an index, market forces determine which indexes become popular.
\textit{Wsoogle} \cite{WSGoogle} is an example of the index approach. It is a global online directory of web services and web service resources. Wsoogle creates a semantic model for each web service operation based on their input and output message definitions and their descriptions defined inside each WSDL document. A similar web service directory is \textit{Woogle}  \cite{dong2004ssw} that looks for similarities between web service operations, exploiting similarity search algorithm.

To avoid potential problems of centralized registries, decentralized approaches aim to achieve flexibility, managing registries in a distributed way and storing the directory information at different network locations.
They can be classified along two orthogonal dimensions: information distribution, that distinguishes between \textit{federation-based} or \textit{P2P-based} approaches, and the use of WS-ontologies, that distinguishes between \textit{semantic-laden} and \textit{semantic-free} approaches.

Federation-based approaches aim to constitute a unique logical registry, distributing information among a set of loosely coupled service registries. In general a federation is created to aggregate registries that have common goals and, consequently, could be interested in the same services.

In P2P-based approaches any node is able to handle the queries it receives, since the absence of a centralized registry that might be a single point of failure. Furthermore, each node may contain its own indexing to some existing services. Finally nodes contacts each other directly, so the information they receive is known to be current. In contrast, in the registry or index approaches there may be significant latency between the time a service is updated and the time in which the updated description is reflected in the registry or index. The reliability provided by the high connectivity of P2P systems comes with performance costs and lack of guarantees of predicting the path of propagation. 

Semantic-laden approaches rely on WS-ontologies mapping techniques like OWL \cite{antoniou2004wol}  or DAML-S \cite{ankolekar:dsw} for incorporating intelligence in the discovery process, for example for cleverly mapping conceptually related terms in queries and advertisements.

Semantic-free techniques, otherwise, are closely related to the traditional service discovery approaches and doesn't adopt semantic languages.
   
Examples of semantic-free federation-based approaches are UDDI and ebXML registries.
UDDI version 3.0 specification \cite{uddi3spec} has been augmented with the support for multiple registries. In particular, it extends the replication and distribution mechanisms offered by the previous versions to support complex and hierarchical topologies of registries, identify services by means of a unique key over different registries, and guarantee the integrity and authenticity of published data by means of digital signatures. This allows various UDDI nodes topologies and several scenarios of possible interaction, such as to enable replication of UDDI entries among the nodes or publishing/subscribing of changes in UDDI entries. Despite these potentialities, the standard only says that different registries can interoperate and the actual interaction policies must be defined by the developers.

To cope this limitation the work in \cite{jagatheesan2003sui} proposed a protocol that enables a query federation for UDDI nodes that are connected in a three-level hierarchy. Queries at the nodes in the bottom level are forwarded to the nodes in the middle and top levels respectively when the queries fail in order to find a result from nodes at the higher levels.
\cite{rompothong2003qfu} enhances the idea of peer-based federation and policy with authentication and authorisation control for access to UDDI entries within the nodes. The federation policy at each UDDI node will specify how to forward queries, and each forwarded node will also authenticate the forwarding node and return query results depending on the authorisation of the forwarding node on its UDDI entries. 

ebXML \cite{ebXML} differs from UDDI on data models and on provided mechanism for the discovery/publishing of web services. 
ebXML is a family of standards based on XML to provide an infrastructure to ease the online exchange of commercial information. Differently from UDDI, ebXML allows for the creation of federations among registries to foster the cooperation among them. The idea is to group registries that share the same commercial interests or are located in the same domain. A federation can been seen a single logical entity: all the elements are replicated on the different registries to shorten the time to discover a service and improve the fault tolerance of the whole federation. Moreover, registries can cooperate by establishing bilateral agreements to allow registries to access data in other registries.
Even if these approaches foster the cooperation among registries, they imply that all registries comply with a single standard and the cooperation needs a set up phase  to manually define the information contributed by each registry. 

METEOR-S \cite{meteor-s} and PYRAMID-S~\cite{pyramids} are the two main representatives of the federation based semantic-laden approaches. They aim to construct a scalable peer-to-peer infrastructures for the publication and discovery of services over private registries. METEOR-S only supports UDDI registries, while PYRAMID-S supports both UDDI and ebXML registries. Both the approaches adopt ontology-based meta-information to allow a set of registries to be federated: each one is ``specialized'' according to one or more categories it is associated with. This means that the publication of a new service requires the meta-information needed to categorize the service within the ontology. This information can be specified manually or it can be inferred semi-automatically by analyzing an annotated version of the WSDL interface of the service. Notice that, if the same node of the ontology is associated with more than one registry, the publication of the services that ``belong'' to that node must be replicated on all the registries.
Services are discovered by means of semantic templates. They give an abstract characterization of the service and are used to query the ontology and identify the registries that contain significant information.
The semantic infrastructure allows for the implementation of different algorithms for the publication and discovery of services, but it also forbids the complete control over the registries. Even if each registry can also be used as a stand-alone component, the selection of the registries that have to contain the description of a service comes from the semantic affinity between the service and the federated registries. For this reason, each node in a METEOR-S or PYRAMID-S federation must accept the publication of a service from any other member of the community. These approaches are a valid solution to the problem of federating registries, but the semantic layer imposes too heavy constraints on publication policies and also on the way federations can evolve dynamically. 

Devised P2P semantic free approaches \cite{li2004psw} \cite{uddibad2} \cite{hu2005app} use Chord overlay for indexing and locating service information. In particular \cite{li2004psw} extracts property-value pairs from service descriptions, that will be used to locate appropriate peers in which at last, the service description metadata will be published.\cite{uddibad2} uses Hilbert Space Filling CUrves for mapping similar Service Descritpions to nearby nodes in the Chord ring. 
In \cite{hu2005app} registry peers are partitioned in numerically ordered subspaces, and each of them maintains links to one peer in each subspace in addition to regular Chord links. Services information is spread embedding semantic information into the peer identifiers, grouping peers by service categories and forming islands on the ring topology.

Other works \cite{wsda} adopts a grid architecture to provide a semitransparent umbrella for distributed data. It does not explicitly focus on web service registry but provides discovery functions for distributed information. Since it only provides data tuple, registry information can be of any format. In particular, for service descriptions this approach proposes WSIL \cite{ballinger2001wsi} containers. It is worth noting that this work does not consider service federations of different registry implementations.

P2P semantic laden \cite{schlosser2002sao} \cite{montebello2003dew} approaches tries to leverage semantic services capabilities to scale to a large numbers of peers, to improve search time optimization and to higher query precision.
These approaches try to solve an orthogonal problem to our, so they don't try to improve and ease registry integration. 

In~\cite{schlosser2002sao} peers with similar services are grouped in concept clusters which are in turn assigned to a specific logical combination of ontology concepts that describes best the peers belonging to the cluster. This ontological partition of network topology enables the network to answer queries consisting of logical combinations of ontology concepts.

In \cite{montebello2003dew} is used an agent based approach to model service requester, service provider and the registry.
Requests are expressed as search by (service provider, service category or service-name) and Inputs/Outputs (required/returned by the web service), and then they are mapped into a DAML+OIL request. In the case of a service provider, DAML+OIL request contains the URL of the service profile that is going to be advertised with the registry. The system is a searching facility that makes use of semantics to retrieve and eventually invoke services. 

As we will see our work is profoundly different from those early described, because we don't have a unique logical registry, physically distributed and  query results depend on the peer to which the query is submitted. Since information at each registry peer depends on the subscriptions done at each peer, in our approach distribution is not transparent to the users.

The most similar approach to ours is VISR (View based integration of heterogeneous web service registry) \cite{WiZNet} \cite{dustdar2006vbi}. This work aims to realize the integration of heterogeneous web service registries and transient web service providers into a common distributed web service registry. To reach this objective it proposes: \textit{communities}, that realize the logical integration of heterogeneous service registries and transient providers depending on common interests; \textit{view}, that add meta information to existing web service registries without changing their existing internal data model;\textit{ profile}, that defines a common abstract service definition, in order to integrate heterogeneous descriptions and to provide the means for the integration of transient providers; a unified means to invoke web services using view as abstraction layer. As our approach, this work tries to solve registry cooperation problems creating an abstraction layer from single registry implementations. The communities are similar to our federations even if they are guided and created from a particular interest, that is revealed from a set of service already published. Our federations are composed by a set of registries that are not obliged to share common interests (a set of similar services). Moreover in our work the information distribution among registries that belong to the same federation is topic based: each element of the federation receives the same information. While VISR views are created starting from the knowledge of services (already published) that belong to various registries, in our approach filter creation doesn't require the knowledge about services currently published in the registries. Our approach doesn't involve stable and transient service provider integration; consequently it doesn't allow to create transient federations. Finally our data model can easily adapt to include information from others data model and it can also easily address problems related to authentication, authorization of federation members. 


\section{Service Model}
\label{sec:model}

The heterogeneity of considered registries and the need for a flexible way to describe available services are the main motivations behind the \textit{DIRE} service model. Exactly it aims to provide to registries of different standards a common way of cooperation, univocally identifying exchanged data.

We observed that different registries tend to provide predefined schemes to organize the information about services. In some cases, they also distinguish between references (stored in registries) and actual contents (put in repositories). 
\textit{DIRE} integrates these elements, to create an abstract information layer that describes the data about services through XML language. We are convinced that the service data representation through XML standards, could be an enabling infrastructure for building, deploying, and discovering of services. Other technologies, such as JAXR~\cite{jaxr} make possible to easily integrate different kinds of registries, providing a uniform and standard Java API for accessing them. However, currently, JAXR specification only includes bindings between the JAXR information model and  registries as ebXML and UDDI v2.0. 

\begin{figure}[htpb]
  \centering{\includegraphics[width=0.5\columnwidth]{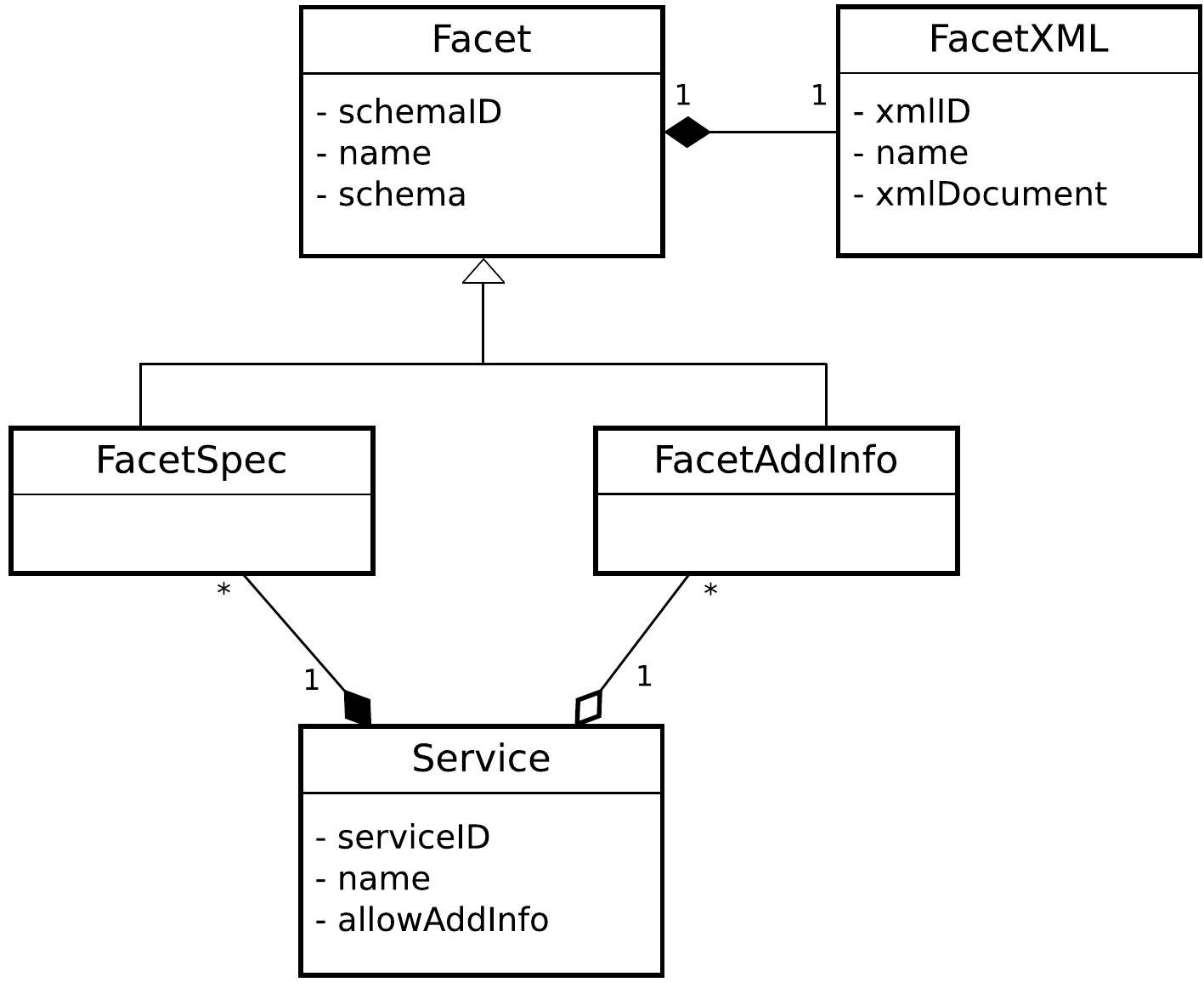}}
  \caption{\textit{DIRE} service model.}
  \label{fig:serviceModel}
\end{figure}

Unlike JAXR, that actually tries to combine ebXML and UDDI, \textit{DIRE} service model, showed in Figure \ref{fig:serviceModel}, is proved to be consistent with any kind of data model adopted in existent registries. The concept of \texttt{Facet}, provided by our model, allows a more detailed and flexible management of the technical information and the compatibility with well-known standards. 

\texttt{Facet}s  characterize a service, describing its features. As we can see in Figure \ref{fig:serviceModel}, each service can be associated to more than one facet (since each Facet represents a perspective from which we depict a service), but one facet refers only to one service. For example, a facet can characterize a service from a functional point of view, containing information about service operations functionality, or it can depict a service from a non-functional perspective containing information about service qualities, such as availability, reputation, response time, etc. 
Facets can contain special-purpose information about elements. For example, we can create \textit{WSDL} Facets to describe the interfaces of a service, \textit{RDF} Facets to add semantics, or \textit{XMI} Facets to specify complex service behaviors through UML diagrams.

Each Facet is associated to a \texttt{FacetXML} element, see Figure \ref{fig:serviceModel}. Facets represent the type of information that describes the service while FacetXMLs are the particular description instantiation, that must comply the type specified by its associated Facet.  In practice a \texttt{Facet} contains an XSD, representing the template that the description has to follow. While a \texttt{FacetXML} contains the XML document that must conform to the schema of the \texttt{Facet}; in other words it contains the actual description compliant to the associated template.
It is important to remark that this mechanism allows each registry to understand service data received from different registries, since each XML description document is transmitted together with its schema. However if the schema contained in a \textit{Facet} doesn't comply with standard or public specifications (such as \textit{WSDL}, \textit{WS-Security}, \textit{WS-Reliability}) it could be necessary an agreement between registries, since each of them should be able to understand the purpose and the semantics of the XML description.

Some kinds of facets contain static information, strongly coupled to the service definition made by the provider. For example the service WSDL is a service description strictly related to the service definition, since it changes during service usage only if the service changes its interface. While other kinds of facets describe features strongly coupled to the service usage, such as service performance perceived by users. For example, this type of facet can represent testing data about the passed test cases that a provider, or eventually, an integrator want to share with possible clients. Moreover it can contain monitoring data, measured during service execution.  
We name the first kind of Facets as specific, see \texttt{FacetSpec} and the second type of facet as additional, see \texttt{FacetAddInfo} in Figure \ref{fig:serviceModel}. 

FacetSpecs describe provided service features and represent the theoretical service characteristics that the service creator guarantees. Since they are strongly coupled with the service they refer to, they can be only attached by the service creator and they must always associated with the service information. 

FacetADDInfos describe the observed service features from the user perspective (provider can also be a user). So they can be attached either by the service creator or by other registries that receive that service. However it is important underline that other registries can publish FacetADDInfos, if and only if the creator allows it, setting the flag \texttt{allowAddInfo} true, (see Figure \ref{fig:serviceModel}).
FacetADDInfos are published and transmitted separately from the service they refer to.

It is worth notice that each update in the service model has to be done in two steps: the deletion of the interested element and the creation of the modified element. 

The distributed setting behind \textit{DIRE} requires that identification and authentication be carefully addressed. Since we can hardly understand the source of exchanged information, we use a digital signature to verify if received messages comply with sent ones and to identify the source of such messages.

In our model anyone can attach a facet to a Service, even if it is not the provider: this feature lets each company use its local registry as a blackboard, and allows a decoupled communication among the different elements of the service-centric system (e.g., runtime monitors might create facets that are then used by the dynamic binder). With \textit{DIRE} we allow to sign and share these facets, allowing the receivers (if they trust the sender) to have a more precise knowledge of the services present in their registry.

\textit{DIRE} can address information confidentiality among federation members. In particular we foresee to manage federation access with an Access Control List (ACL), in which accounts and credentials of the trusted registries are stored. The information exchanged within a federation should be expressly encrypted with asymmetric encryption techniques.


\section{Motivating example}
\label{sec:caseStudy}
This Section introduce a running example that will be used in the remainder of the paper. At this stage it has the objective to illustrate a possible application of our work in a real-world scenario and to exemplify how data model is exploited in different contexts. In the following sections we will leverage this example to clarify how information exchange takes place and to measure other interesting properties.  

\begin{figure}[ht]
  \centering
  \includegraphics[width=.5\columnwidth]{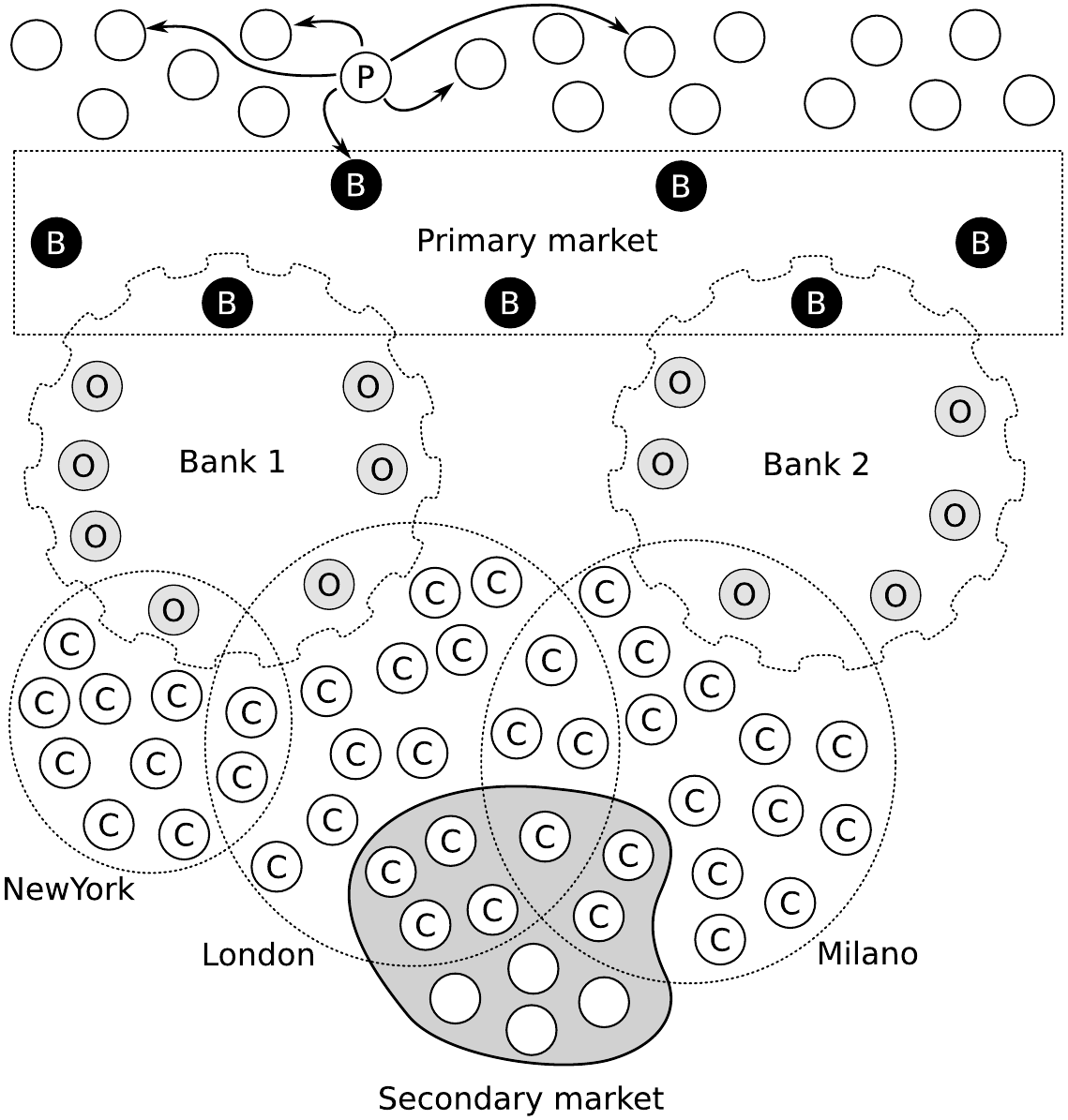}
  \caption{Case study}
  \label{fig:scenario}
\end{figure}

Our example is shown in Figure~\ref{fig:scenario} and is set in a financial environment. There are the following actors: service providers companies, banks and clients.

A service provider (\textit{P} in Figure \ref{fig:scenario}) offers a service to compute the salaries of its employees and an information brokering service, to produce, for example, domain studies or market forecasts. The service provider company wants to keep the first service private, while it wants to disseminate the second: if other entities discover the availability of such a service, the company increases its revenues.

The gear shapes depicted in Figure \ref{fig:scenario} represent commercial banks. Each of them is composed by a central bank that controls a set of affiliated subsidiary offices (respectively \textit{B} and \textit{O} in Figure \ref{fig:scenario}).

The central bank also cooperates with other central banks of different national companies and the national secretary of the treasury in the primary capital market, in which takes place the issue of financial, bonds, etc. that will be exchanged in the secondary capital market.
Entities involved in the primary capital market, even if they interact in different scenarios and communicate with other entities in different environments, can constitute a very small and stable federation.

A real-world scenario can take place when entities involved in the primary capital market need to evaluate subscription offers made by financial operators. In particular they need to determine the convenience of a financial operation with an external enterprise that operate in a specific sector, analyzing information about the enterprise and the market. Hence central banks can be interested in information brokering services that produce financial studies, analyzing information about particular enterprises (such as balance sheet, company type, credentials), the market trends (where and how that company is placed in the national capital market) and the companies that do the same commercial activities in that country or in the world. Since central banks share common interests, if one of them is using the information brokering service, it might want to share with other federated central banks performance information about the adopted service, through, for example, the topic \textit{PrimaryMarket}. Doing this allows central banks to create a common knowledge between their registries and to select within a group of services with similar functionality, those that have the best performance.

The central bank has to coordinate and control its subsidiary offices, in order to give them the same directives and share common objectives that belong to the company. Therefore the central bank needs to establish a common way through which share contents and information with subsidiary banks. Hence each central bank could share public adopted services with its subsidiary offices, creating the federation with the topic \textit{CommercialBank}. 

Each subsidiary office, in general, has some clients (\textit{C}  in Figure \ref{fig:scenario}) that address the bank to do several operations, such as open an account, get a loan or invest. Clients could be not only physical persons but other enterprises. Each client can address more than one subsidiary office, not obligatorily belonging to the same commercial bank, for example when he has more than one account in different banks or he can deposit/withdraw money from several subsidiary offices of the same company.

Clients can share common interests, either since they address the same subsidiary office or have a similar financial profile. The first group of customers and their subsidiary bank, clustered in circle shapes shown in Figure \ref{fig:scenario}, would know information about changes in the interest rate provided by reference banks. 
While the second group of costumers, clustered in the shaded shape, would know information about financial market trend. Their common interests make convenient that clients group each other to constitute a federation.
 Hence clients could be interested in the information brokering services, that can collect information about the convenience to deposit money in a specific bank or of an investment. 
 For example in a first scenario a client involved in the shaded federation is using the information brokering service, observing good performance and results. Consequently he wants to propagate the use of this service to other federated clients. So federation exploits the group \textit{``SecondaryMarket''} to propagate the information about the service of interest. The federated customers become aware of the new service, retrieve its data, and store them in their registries. Finally the information brokering service is now available to the whole set of registries of the federation.

\section{Delivery Manager}
\label{sec:deliveryManager}
This section introduces the main elements of \textit{DIRE} (DIstributed REgistry), which is our proposal to support loose interactions among registries. 

Most of other approaches -presented in Section~\ref{sec:relatedWork}- build up an unique logical registry, internally composed of several distributed nodes. Those approaches use a distributed algorithms to solve typical problems and bottlenecks of centralised registries, such as the unique point of failure, the high response time, and the requirement of high availability of the information. Anyway, these works shares the information overloading problem with large-centralised repositories, that lead to a low-precision discovery process.

In order to solve these problems, our approach introduces a different cooperation style among the involved parties. We propose to create three different zones: a \textit{private registry}, \textit{trusted federation network}, and a \textit{public marketplace}.

The first of these zones, is the \textit{private registry} of each user. Our proposal foresees that each party (i.e. service providers and all potential clients) has his own local registry, on which it has (and wants to keep) the entire control. This proposal is in line with current enterprise architectures, that prescribe the usage of a local registry, able to maintain information regarding useful services. The player should also have is own local SOA environment, that provides also other basic features such as process execution, monitoring, and recovery. The registry is able to glue these parts together, enabling a blackboard communication style. Each component can add information to the registry, or look for services having particular characteristics. For example, the monitoring part of the customer's enterprise architecture is able to measure performances of services used, detecting failures, and modifying the corresponding information stored in the local registry. Moreover, the recovery planner should leverage the enriched service information present in the registry to better select the more suitable service. Obviously, this cooperation style is applicable only on a small, controlled environment.
It is worth to say that each organisation inserts explicitly interesting services in his private registry, performing an important service selection. This operation avoid the information overloading problem, ensuring that each service discovery returns services with higher quality, since it is applied on a reduced and controlled domain. In this way it is possible to enact organisation's peculiar policy simply controlling this service selection. For example, suppose that an organisation have particular agreements with an hotel chain, and it wants to ensure that his ``business trip organiser'' process selects only these hotels. In order to ensure this policy, the organisation can manage accurately the insertion of hotel services in his private registry. Subsequent hotel service discoveries will find out only booking services of the preselected chain. For this reason, the discovery process have an higher precision (i.e., almost only relevant services are found), but potentially with lower recall (i.e., not all relevant services are found). In order to higher the recall, our proposal introduces a ``\textit{public marketplace}'' and the ``\textit{federations}''.

Service providers can publish their services to the \textit{public marketplace}, allowing interested customers to discover their services. Like in the ``real'' marketplaces, providers broadcast\footnote{The name of the project ``dire'' is also an Italian verb that means \textit{to speak}, that underline this cooperation style.} qualities of their products, and customers listen to interesting offers using simple matchmaking filters. In this phase there is a loosely-coupled cooperation mechanism, that make possible to cope with very-large communities. 
When a customer receives a particular service description, he can analyse it better and decide whether it matches with his requirements or not. If it correspond to the customer's expectations, the service will be inserted into the registry. It is worth to notice that with this action, the service is immediately inserted into the customer's applications, since it becomes  visible to the registry's client.

Different cooperating organizations may desire to share some services, able to fulfill some common requirements. For this reason, organizations can be grouped in \textit{federations} to allow for the redistribution of interesting information among all partners. Obliviously, one organisation can belong to several federations.
Each member can declare that a service present in his local registry is pertinent to the whole federation, and \textit{promote} that service into the federation. The effect of this operation is that the information regarding that service is spread among the federation, and is inserted into the local registry of each member. In this case the service selection is performed by the promoting peer, and other nodes trust him, accepting the services he promotes. This style of cooperation requires that members of the federation trust each other accepting the services promoted by each member.

\begin{figure}[ht]
  \centering
  \includegraphics[width=.6\columnwidth]{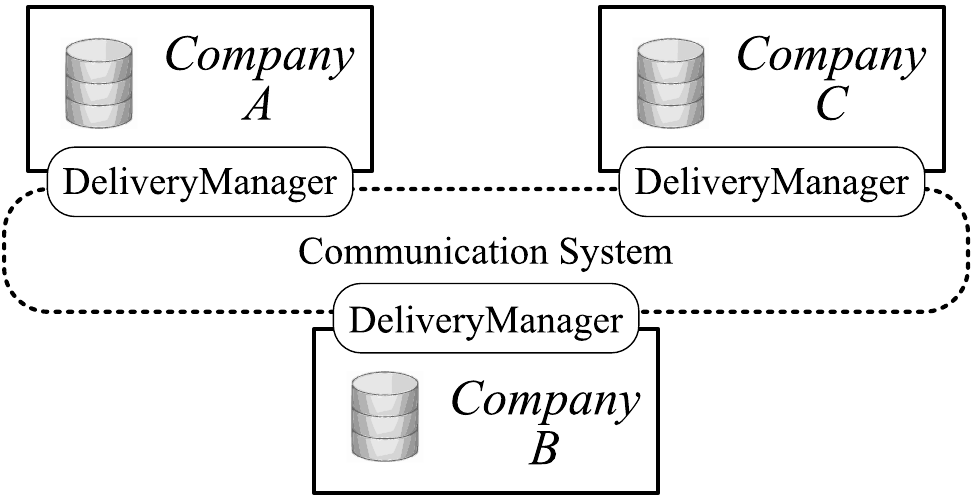}
  \caption{Distributed architecture of the DIRE}
  \label{fig:direArch}
\end{figure}

DIRE aims at supporting these kind of cooperation among different proprietary registries by means of two elements: a global \textit{communication system} and a \textit{delivery manager} associated with each registry, as shown in Figure~\ref{fig:direArch}. 

The former allows message exchanging among different delivery managers, in a peer-to-peer manner. The communication system act as a common reference for each delivery manager, and provides functionality for managing both the marketplace and federations.
The core of this communication system is based on REDS~\cite{reds}, a distributed \textit{publish~/~subscribe}\footnote{The reader that needs an introduction on publish~/~subscribe systems, may read the Appendix.}~\cite{DistributedPS}, with advanced features such as reply support~\cite{PubSubWithRep}, self-healing~\cite{frey:phd}, and self-optimising~\cite{migliavacca2007aps} capabilities.
In \textit{publish/subscribe} systems, components do not communicate directly, but rather communication is mediated by a \textit{dispatcher}. Components define active agents, which send events to the dispatcher (\textit{publish}) and decide the events they want to listen to (\textit{subscribe}/\textit{unsubscribe}). The dispatcher forwards (\textit{notifies}) events to all registered components, which react to them.
Distributed publish/subscribe systems are able to split the dispatcher among several nodes, guaranteeing the logical integrity. In this way it is possible to create a scalable middleware, able to manage very large networks. Our approach leverage this feature proposing to create a large communication system, able to connect each different delivery manager.
The reply mechanism allows us to adopt a query-response mechanism: nodes looking for an information publish a query message, that will be delivered to nodes that are able to manage that query (i.e. they had subscribed for that kind of query); these nodes can reply to requester with a message containing the response of the query. 
Distributed publish/subscribe middleware can adjust their own internal structure to react to node failures, or to optimise the overall performance, ensuring a reliable and efficient communication system; our proposal indirectly uses those results to guarantee fault-tolerance and scalability.

The delivery manager acts as \textit{facade}, allowing personnel of organisations to manage the \textit{marketplace} and the \textit{federations}. The delivery manager is connected with the organisation's local registry and the global communication system, and it is able to support the information flow in the two directions. For this reason, the delivery manager is able to perform the adequate conversion between the service model used by the registry and the one presented in Section~\ref{sec:model}.
When it is requested to expose a service in the marketplace or to promote the service in a federation, the delivery manager accesses the local registry, convert the retrieved information into a set of messages, and propagate them with the given mechanism. Dually, when one is searching for interesting services exposed by other providers or retrieves information promoted by other trusted parties, it is able to convert the received set of messages into registry's processable information, and store it in the local registry. Moreover, the delivery manager offers also management functionality, allowing one to create, join, or leave a federation, and manage the connection with the communication system.

\begin{figure}[ht]
  \centering
  \includegraphics[width=0.8\columnwidth]{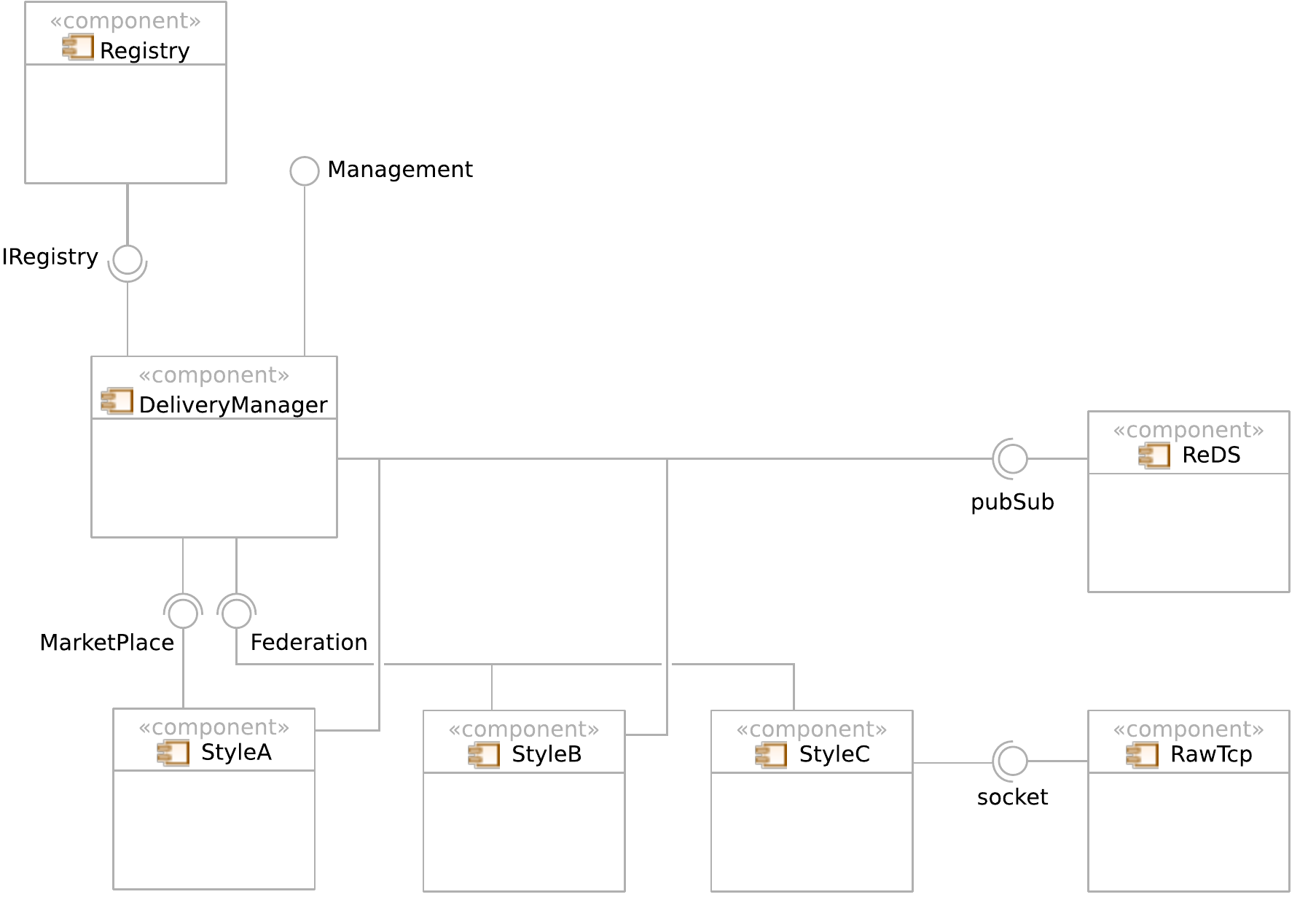}
  \caption{Architecture of the Delivery Manager}
  \label{fig:deliveryManagerArch}
\end{figure}
Figure~\ref{fig:deliveryManagerArch} details the architecture of a single delivery manager, showing how it manages the interaction between the registry and the communication system.
On the upper side of the diagram it is possible to notice the registry, used by the delivery manager to store or retrieve service descriptions. On the right side there are the connections with the global communication system (\texttt{ReDS} in the diagram).
As stated before, an organisation must interact explicitly with the delivery manager, specifying  operations to carry out both on the marketplace and on federations; for this reason, the delivery manager exposes the \texttt{management} interface, allowing one to share services, express its interests, join/leave federations, and promote services.
Designing the local architecture, we chose to keep separate the concepts of marketplace and federations, from their particular implementations (i.e., the protocol for information spreading). The former are managed directly by the delivery manager itself, demanding to \textit{cooperation styles} the real information exchanging. Our proposal comes with pre-defined cooperation styles, but we allow one to design its own cooperation style, compliant with our definition of marketplace or federation, and plug it in the delivery manager. In the diagram, \texttt{StyleA} is a marketplace cooperation manager, while \texttt{StyleB} and \texttt{StyleC} are compliant with the federation cooperation pattern.
It may happen that, given a particular cooperation style, there are communication systems that performs better than REDS. For example, some protocols requires a point-to-point communication among the parties; in those case a simple \textit{TCP socket} performs better than a publish~/~subscribe system, which introduces an high overhead. For this reason, each cooperation style may use a third party communication system, as \texttt{StyleC} does with \texttt{RawTcp}.


The remaining part of this Section will dig down in the delivery manager, analysing furthermore the public marketplace, and the federation-based cooperation, and proposing specific \textit{cooperation managers}.

\subsection{Public marketplace}
\label{sec:marketplace}

The delivery manager allows one organisation to share some elements (services), and to declare its interests in particular kinds of elements. This way, it is possible to transfer some service information from the provider registry to the consumer registry.
Analysing better the marketplace cooperation, it is possible to notice two main operation: a service provider may \textbf{share} one of his services, and a customer may \textbf{declare interests} on services satisfying particular conditions. The marketplace implementation should allow one to perform both these operations, guaranteeing that if a shared service information satisfies a customer's requirement, that information is delivered to him.

Services can be \textit{shared} by providing the delivery manager with what should be shared. The company can choose to share an information that it creates on its registry (i.e., it is not received from another registry). Thus the manager retrieves this information from the registry, transforms it into the right format, and publishes it using the selected cooperation style.
It is worth to say that different companies can share different kind of information regarding the same system. The system grant the service provider the right to share the service, with the set of \textit{facet specification}. Other companies may receive that service, measure their experience with that service, and maybe share this experience with other customers. For this reason, each customer can share an \textit{additional information facet}. 
An eventual specification facet added to a shared services, requires to re-send the whole service description. The delivery manager ensures that only the node that originally shared the service can add specification facets. Other nodes can only describe additional properties on the service and, if allowed by the service provider, they can share them as \textit{additional information facets}. 

For example, a service provider may decide to share a service for getting trades of stocks. For this reason, that company will create its service description, with facets such as WSDL interface, Quality of Service, and pricing information. Other customers that somehow have a reference to that service, may share their additional information facets, containing for example their monitored performance, the test they made, or their rating of that service.

A customer can \textit{declare its interest} on a service description, using different level of information: he may know in advance the service he wants, and ask the delivery manager to retrieve it, or he should specify its requirements in a formal way, asking the delivery manager to forward him matching services, or finally he may desire to enrich his view on a service, retrieving additional information created by other customers. 

Because of commercial agreements between the parties, the client may know in advance the service it wants. In this case, the selection can be precise: the unique identifiers of interested elements are used by the delivery manager to retrieve exactly the specified element.

Otherwise, an organisation can select interesting services specifying some constraints that the service description must hold. For this reason, our solution leverages the service model presented in Section~\ref{sec:model}, and allows one to express a conjunctive constraint on the \textit{facet specifications}, such as: $ c_1 \wedge c_2 \wedge c_3 \wedge ... \wedge c_n $. Each sub-constraint $ c_i $ is able to check whether a \textit{facet specification} satisfies some properties, expressed using the \texttt{XPath} language. Moreover, it is possible to specify the type of information for each sub-constraint, expressing the \texttt{XML Schema} document of the facets. The overall service description matches with user requirements if the constraint is satisfied, i.e., each sub-constraint $ c_i $ is satisfied.\\
For example, if an organisation is interested on financial services, able to retrieve the last trade of a stock, with a response time lower than 100ms. This requirement is a conjunction of two simpler demands: 
\begin{itemize}
\item the former is on the interface of the service; so the XML Schema is set to the \textit{WSDL} one. The service must have a method called \texttt{getLastTrade}, so the corresponding XPath is\\ \texttt{//operation[@name='getLastTrade'] }.
\item the latter is on its QoS, so the XML Schema is set to the adequate one, and the XPath is\\ \texttt{/QoS/response[case='worst']/time[@format='ms'] < 100 }.
\end{itemize}
When a customer decides to declare its interest using a filter like this, the delivery manager ensures that matching services will be delivered to him. 

Moreover, an organisation may desire to express interest also on the information provided by other clients (i.e., on \textit{additional information facets} of our service model). For this reason, it is possible to declare an interest on additional information facets, specifying both the id of the referring service, the type of the interesting facets (with an \texttt{XML Schema}), and an \texttt{XPath} expression containing the constraint that the facet must satisfy.\\
For example, suppose that one would like to faster the testing phase of a service, collecting significant tests realised by other customers. In this case, he should declare an interest on additional information facets with type \textit{SoapTest}, XPath \texttt{/SoapTest[count(testcase) > 10]}, and that refers to the previously matched service.

The marketplace cooperation style requires some common reference point, that delivery managers can use to exchange service details. Currently our prototype leverages the global communication system, that acts in this case as the required reference point, making able different delivery managers to exchange messages.

\subsubsection{Publish~/~Subscribe }
\label{sec:marketplacePubSub}

The delivery manager have a cooperation manager compatible with the marketplace, that is based on the publish~/~subscribe paradigm. In order to describe the behaviour of this mechanism, we present how a sharing is managed, and in which way it is possible to declare a interest.

In our proposal, when one decides to \textit{share} something, it is performed a publication of a message containing the selected element. For this purpose, we'd created two main messages: one able to host services, and the other one for additional information. When a service provider, such as the \textit{InformationBroker} of our case study, wants to share one of its services, the delivery manager picks up the service itself, and the set of facet specifications he created for that service. In this way, a message is sent through the publish~/~subscribe network, allowing user to retrieve interesting services. In a similar way, users of the InformationBroker's service may share tier additional facets, containing for example, the executed \textit{testCase}s.

On the other side, when one expresses its \textit{interest}, it is performed a subscription on the publish~/~subscribe middleware. These subscriptions will be matched against shared elements, being able to forward elements to interested users.
According with types of interests, we may have subscriptions on service descriptions or subscriptions on additional properties.
If the user is searching for a service, he will generate an interest that will matched against descriptions set by the service providers. In particular, he may specify either the desired \textit{serviceId} or a set of \textit{XPath} that service description must match. A message matches this subscription if it contains a service description that, respectively, either has the specified id or its facet specification are compliant with the given XPath constraints. 
The other type of subscription is generated from an interest on an additional descriptions on a  service. In this case, the filter is only matched against messages containing additional information, and verify whether it refers to the right service, and if its content satisfy the XPath condition set by the user.

In this way, the middleware ensure that messages containing service descriptions are forwarded to interested customers. The solution, as presented so far, has two main problems: it cannot guarantee a distributed coherence, and doesn't allow new customers to retrieve shared services.
The first problem is due to the unreliable network we base on: since nodes can crash, have temporary failures, or network problems, it is possible that some messages are lost. Anyway, if there are two customers having the same interest, it is possible that the network delivers a matching service only to one of these nodes, since the message headed to the other node is lost. If the deletion of a service is achieved through a message, the problem can be even more emphasised, since a customer may have in its registry a service that doesn't exist.
The second problem is experienced whenever a customer express an interest that an already shared service matches. Since the message containing the service description is already been sent, the subscription performed by the customer will never match previously-shared services.

In order to solve these problems, the propagation of service information with this method is subject to \textit{lease} contracts, a typical concept of many distributed systems (e.g., Jini~\cite{jini}). When the lease expires, the information is not considered valid anymore and it can be deleted; only a \textit{renew}, which requires that the information be retransmitted, allows for extending the validity of such information. The delivery manager can perform automatically this operation, so it is guaranteed that when the information is re-sent, all interested registries can retrieve it.

\subsection{Federations of trust}
\label{sec:federation}
A federation groups together different organisations according to some common interests or goals. A federation of delivery managers reflects some kind of real-world federation or common interest among their owning organisations. For example, in our scenario a \textit{Bank} is a federation of several offices (either central or subsidiary). A financial service may be promoted into this federation, so facets describing it will be present in the registry of each member belonging to the federation, so it can be discovered easily and automatically.

Federations, in our proposal, are created explicitly by a peer, which manages their life-cycle. This node is also present in many real-world federations: for example, the central office of a bank takes care of the overall bank management, comprehending also the correspondent federation. 
Each partner of a federation should be able to correctly join a federation using the information set by the federation creator. After its join, he becomes able to promote services or retrieve services promoted by other members. For example, a subsidiary office should be able to correctly join the federation representing its Bank, being able to receive bank's specific services.

\begin{figure}[ht]%
  \centering
  \subfigure[Creation]{\label{fig:fedDir:creation}\includegraphics[width=.3\columnwidth]{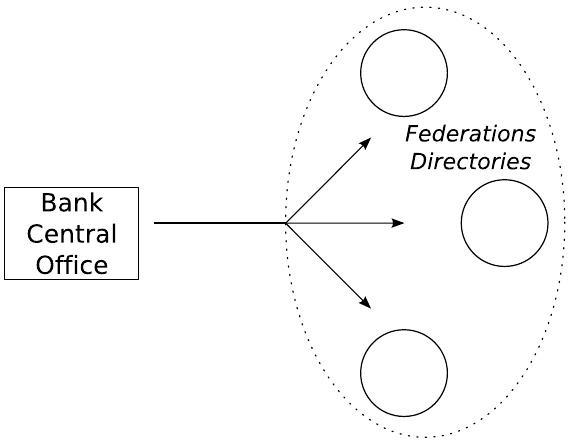}}\qquad
  \subfigure[Directory discovery request]{\label{fig:fedDir:dirReq}\includegraphics[width=.3\columnwidth]{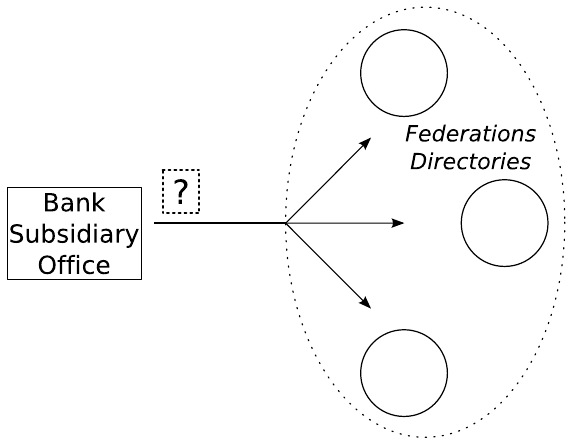}}\\
  \subfigure[Directory discovery reply]{\label{fig:fedDir:dirRep}\includegraphics[width=.3\columnwidth]{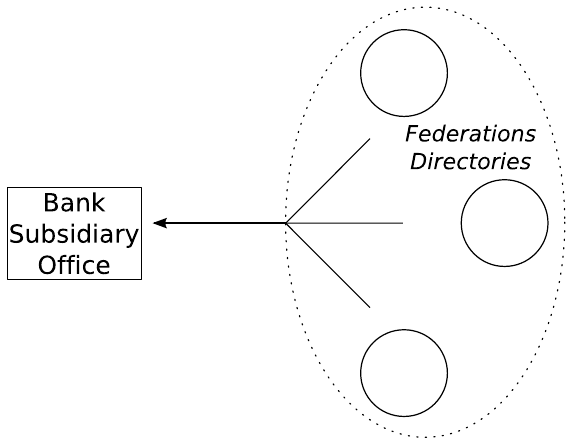}}\qquad
  \subfigure[Directory usage]{\label{fig:fedDir:dirUsage}\includegraphics[width=.3\columnwidth]{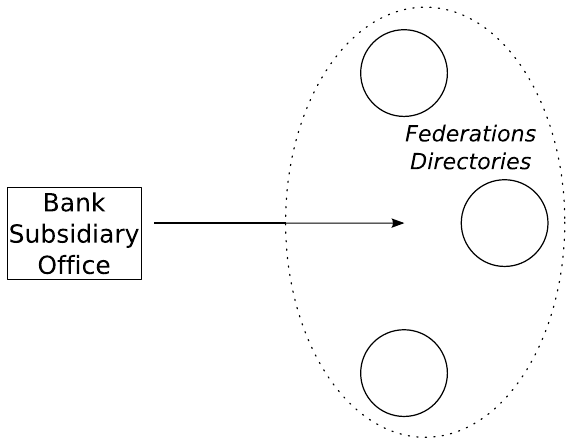}}
  \caption{Creation and join of a federation.}
  \label{fig:federationDirectoryActivities}
\end{figure}

The flexible architecture of the delivery manager --presented in Figure~\ref{fig:deliveryManagerArch}-- makes possible to keep separate the concept of a federation, managed by the deliveryManager itself, from the protocols able to implement it, enforced by cooperation styles. In order to support the communication within a federation, it is required that all members select the same cooperation style (i.e., the same protocol for exchanging messages). For this reason, our proposal requires that when a federation is created, a precise style is chosen. All peers, in order to join the same federation, must be able to understand which is the selected protocol, and initialise the corresponding cooperation manager to join the federation.
For this reason, it is required that the manager of a federation is able to pass federation details to possible members, allowing them to have the adequate amount of information to join it. To facilitate these operations, we have created the \textit{FederationDirectory}, a component able to host information regarding federations. Both federation creators and aspirant members can contact this component, and exchange necessary information. 
Our proposal adopts a decentralised architecture, that uses the global communication system to receive the events for managing information regarding a federation. Each instance of the federation directory is subscribed to the topic \texttt{FederationDirectory}. When one needs to contact a federationDirectory, it can send a message to this topic. Figure~\ref{fig:federationDirectoryActivities} details operations performed using this mechanism.

Figure~\ref{fig:fedDir:creation} shows the creation of a federation representing a new bank, performed by its central office; this operation implies that the central office's delivery manager sends a notification to all the federation directories using the adequate topic. The information exchanged is subject to a lease contract, that ensure the distribute coherence of the system. This means that the creator of a federation periodically have to renew information regarding its federation, allowing new federation directory to retrieve it. 
In some situations partners decide to end their cooperation, so the federation is dismissed. The manager of the federation has to apply this decision also informing the delivery manager of all partners and removing the entry stored in the directory service. Since federation information is subject to a lease, if the manager does not renew it, the federation is automatically dismissed. It is required that all partners periodically check if the federation is still present in the federation directory, so they can discover if it is still active. Nevertheless, the federation directory can be asked to remove information on a particular federation, before the expiration of the lease. This allows the organisation that is managing a federation to quickly dismiss a federation. This forced dismissal is accomplished through REDS messaging using the \texttt{FederationDirectory} topic. If the manager needs a faster federation dismissal, it can use the communication protocol and deliver an adequate dismissing message to all partners. This method is faster than the previous one, but it is also less reliable: temporary failures or incomplete delivery of the dismissing message may make a federation member miss the notification. For this reason, we use a combined approach: the manager uses the communication protocol to send a message and dismiss the federation, but all partners periodically check if the federation they joined is still present in the federation directory.
The distributed architecture of the whole system requires that a delivery manager can automatically discover and select a federation directory, so it can use the exposed service and fetch the list of federations. For this reason we propose to use the reply functionality of the chosen publish/subscribe middleware. A delivery manager can send a directory discovery request on the topic \texttt{FederationDirectory}, as shown in Figure~\ref{fig:fedDir:dirReq}. Each available public directory should respond using the reply mechanism implemented in REDS, and specifying in the message the URL of its endpoint (Figure~\ref{fig:fedDir:dirRep}). In this way the delivery manager can collect transparently the list of available directory services, select one of them and use its functionality (Figure~\ref{fig:fedDir:dirUsage}). 

Considering the operations that a federation-compliant cooperation style must provide, it is possible to notice three basic operations:
\begin{itemize}
\item \textbf{creation}: Each style must be able to initialise the federation, allowing other peers to join and exchange messages. In this step, it is required to declare all properties of the federation, including its name, the chosen cooperation style, and the information for the connection.
\item \textbf{join}: Each style must be able to connect to an existing federation, using the information set by the federation creator. After a successful join, the node should be able to exchange messages with other members.
\item \textbf{promotion}: Each member of a federation should be able to promote services to its federation. Dually, each member of a federation should be able to retrieve promoted services, inserting them into its private registry. For example, the central office of the bank should be able to promote services to all subsidiary offices; these one should receive the promoted service, inserting it in their private registry.
\end{itemize}

The remaining part of this section presents three federation managers, comparing their characteristics in terms of level of coupling, reliability and number of messages exchanged.

\subsubsection{Publish / Subscribe}
\label{sec:pubsub}

A first way to establish a communication system among all the different parties of a federation is to use the global publish/subscribe middleware. In this way, the delivery manager treats federations as special-purpose subscriptions, using topic-based filters. The federation creator generates the topic (i.e., the identifier of the federation), and insert it in the federation directory. 
When other peers join a federation, they can retrieve the topic associated with the federation from the \textit{federationDirectory}, and subscribe themselves to it. This ensures that every time there is a message for that topic, it is received by all the participants of the federation, creating a multicast communication group.

When one decides to promote a service in a federation, it is possible to leverage this multicast communication group to spread this information. In this way, all members will receive a message containing the promoted element, allowing them to insert it into their private registry.

In order to ensure a distributed coherence, allowing also new members to receive previous promotions, our proposal prescribes that the propagation of promotions is subject to a \textit{lease} contract. When the lease expires, the information is not considered valid anymore and it can be deleted; only a renew, which requires that the information be retransmitted, allows for extending the validity of such information. The delivery manager can perform automatically this operation, so it is guaranteed that when the information is re-sent, all interested registries can retrieve it. 

The characteristics of this communication style, suggest that it should be used whenever different members are highly decoupled, the federation is small (considering either the number of service promoted or the numbers of members), or there are an high turnover of federation members.

The high-level of decoupling between each member, requires us to manage large federations that cross-cut several networks. Using the lease mechanism, we are able to tolerate transient network failures, ensuring high reliability. Moreover, we can use a renew rate faster than the expiration rate, setting for example the former to one day and the latter to one week. In this way, we ensure that transient (i.e., that last less than one week) network failures can be tolerated, at the cost of slower promotion deletion (i.e., once a promotion is sent, it last at least one week).

Analysing the number of exchanged messages, it is possible to state that each single element promoted is sent periodically to each node of the federation, and this promotion is renewed periodically. The network traffic, in terms of network messages, is deterministic, and is equal to: 
$$ P \cdot (N - 1) \cdot {D \over T_{renew}} $$
where \textit{P} is the number of promotions, \textit{N} is the number of peers, \textit{D} is the duration of the federation, and \textit{$T_{renew}$} is the renew period.
It is worth to say that the network traffic depend directly on the number of elements promoted, and on the number of members of the federation, but doesn't require any setup cost for new members.

Analysing our running example, this style of federation fits well both for the \textit{primary market} and for the \textit{secondary market}. Members belonging these federations  to different organisations, so it is impossible to ensure a reliable network. Moreover, the federation is relatively slow, since the number of central offices is tight.

\subsubsection{Publish / Subscribe with Reply}
\label{sec:pubsubreply}

A more sophisticated way to exchange promotion in a federation is based on request~/~reply communication schemas. For this reason, it is possible to leverage the reply support~\cite{PubSubWithRep} of the communication bus to create more coupled federations. The resulting approach is more efficient (i.e., send less messages and new members receive earlier pre-existent promotions), but do not tolerate even transient network failures. For these reasons, it should be used only within controlled network, for example to connect different subsidiary offices of the same corporation.

Like the previous publish~/~subscribe mechanism, the creator of the federation has to select the topic, that will be inserted in the federationDirectory. In order to join, all members have to subscribe to this topic, becoming able to send and receive messages on it. 

Promotions are achieved by sending a message containing the promoted service in this topic. Differently from the previous method, promotions are not subject of lease contract, and last until they are explicitly discarded. Supposing that federations are stable (i.e., new members are less frequent that new promotions), the number of messages sent is lower.
When one joins a federation, it is possible to leverage the request~/~reply communication schema to retrieve pre-existing promotions. 
For this reason, the new member should send a request on the topic of the federation, requiring that all partners replies with services they promoted.

This cooperation style is more efficient than others, but imposes stronger network requirement, otherwise the overall reliability is not guaranteed. The efficiency of the network can be measured analysing the number of messages required to spread promotions in a federation. 
Calling \textit{N} the number of members of the federation, for promoting an element, the network sends $ N-1 $ messages, one for each receiving member. 
Considering also new members, they send a request message, and then receive all promoted services. The size of request message is strictly lower than messages containing promoted elements, so it is possible to drop them, and state that for each new member it is required to exchange approximately $ P $ messages, one for each promoted element. 
The overall number of messages is:
$$ P \cdot (N-1) $$
where \textit{P} is the number of promoted elements, and \textit{N} is the size of the federation, measured in number of peers.

Considering our running example, it is possible to use this cooperation style to exchange services within a single bank. Managing accurately the configuration of the communication system, it is possible to ensure that all offices of the same bank are connected using the internal network, ensuring in this way an high availability of the network, and a corresponding high dependability of the federation. The result of those more stringent requirements on the network is paid in terms of better performances.

\subsubsection{Gossip}
\label{sec:gossip}
Gossip protocols~\cite{ganesh2003ppm} are based on peer-to-peer communication between members, inspired by theory of epidemics. In these algorithms, nodes with information to disseminate are ``infected'', and may contaminate other peers with a certain probability. The overall goal is to infect as many nodes as possible (i.e., spread the information). These protocols ensures high level of dissemination, even if the underlying network is unreliable, requiring only point-to-point message exchange. For this reason, we decided to create a separate communication bus using raw \textit{tcp} sockets, without using the publish~/~subscribe global communication bus. 

We propose to use those algorithms to spread promotions of services. To promote a service, a peer contacts some members of the federation to notify them the promotion of the service. The notified members may contact other members, and so on. The gossip protocol does not guarantee that all federation members receive a given message. Instead, there is a high probability that most members receive a given message, a low probability that only a few members receive a given message, and a vanishingly small probability that an intermediate number of members receive a given message. Simulations of gossip-based federations showed that more than 98\% of the members are reached by a given message, on average.

The creator of the federation should publish the initial contact of the federation, that subsequent members can use to successfully join the federation. 
Joining a federation that uses the gossip strategy requires knowing a federation member. Peers that want to join the federation have to contact this known member, to acquire information on the rest of the group. The gossip protocol requires every member know a subset of federation members. Each node can unsubscribe from the federation by informing known members that it is leaving the federation.

Services in gossip-based federations are only published once(i.e., they are not subject to lease contracts), letting the gossip protocol to spread it. When a peer joins the federation, it asks another member for the list of promoted services.
The nature of gossip protocols requires that deletion are managed adequately. In particular, it is not enough to delete a service, because it can be gathered from other nodes --aware of that deletion--, which are trying to spread that information. For this reason, a promotion deletion is an information that each peer must keep, and spread like service promotion. The drawback are only that the deletion of promotions requires a certain traffic on the communication bus, and each peer have to keep a deletion list. 

The traffic of a network using gossip algorithms is difficult to evaluate and an exact value cannot be provided because of the non-deterministic nature of the protocol. Since the gossip strategy uses the SCAMP protocol, the average number of network messages exchanged to promote services is
$$ P \cdot N \cdot log(N) \cdot (C + 1)$$ 
where \textit{N} is the number of nodes in the network, \textit{C} is a fault tolerance non-dimensional parameter, whose typical value is 2, and \textit{P} is the number of promoted elements.
The formula above shows that the traffic for service promotion is independent of the running time of the federation. 
In addition to service promotion traffic, the gossip strategy generates network traffic to manage group knowledge. Even if the messages exchanged for this purpose are smaller, in long-running federations with a low promotion rate, they can require a cost comparable to the service dissemination. Most of this traffic is generated by periodic heartbeats and resubscriptions.  The number of heartbeat messages is approximately
$$ N  \cdot log(N) \cdot (C + 1)\cdot { D \over T_{heartbeat}} $$ 
where \textit{N} and \textit{C} keep their previous meaning, \textit{D} is the interval of time in which the federation is kept running, and \textit{$T_{heartbeat}$} is the heartbeat period, whose typical value is 1 day.
The number of messages for resubscription is, on average, approximately equal to
$$ N \cdot log(N)^2 \cdot (C + 1)^2  \cdot {D \over T_{resubscription}} $$ 
where \textit{N}, \textit{D}, and \textit{C} keep their previous meaning, and \textit{$T_{resubscription}$} is the resubscription period, whose typical value is 1 week.

Those characteristics suggest the usage of this communication style on large and decoupled federations. Gossip protocols ensures low management costs even with unreliable networks, guaranteeing a good dissemination rate (i.e., on average there is only 2\% of lost messages). Looking at the proposed case study, we propose to use this kind of federations to manage interaction among a bank office and its clients.

\subsubsection{Comparison}
\label{sec:comparison}

Upon creating a federation, the managing organisation must choose the communication strategy that is used for service promotion and deletion. This Section provides a summary comparison among the proposed communication strategies, in order to help one in choosing the one that best fits his requirements.

This comparison is performed considering several dimensions:
\begin{itemize}
\item \textit{Number of messages per promotion}: is the number of messages sent on the underlying network for each promotion made.
\item \textit{Number of messages for maintenance}: is the number of messages sent for the maintenance of the communication network. In federations with few promotions, this traffic can represent the dominant factor.
\item \textit{New member time}: is the time required by the protocol to ensure that a new member receives promoted elements.
\item \textit{Failure sensibility}: measures how the protocol is sensible to network failures.
\item \textit{Coupling}: measures how a given protocol tend to couple different parties. 
\end{itemize}

\begin{table}[ht]
  \centering
  \caption{Comparison}
  \label{tab:fedComparison}
  \begin{tabular}[ht]{r|c|c|c}
                                    & \textbf{PS} & \textbf{PS-R} & \textbf{Gossip} \\ \hline
    \textit{\# msg per promotion}   &  High       & Low           & Medium-Low      \\
    \textit{\# msg for maintenance} &  None       & None          & Medium          \\
    \textit{New members time}       &  High       & Low           & Low             \\
    \textit{Failure sensibility}    &  Low        & High          & Medium          \\
    \textit{Coupling}               &  Low        & High          & Medium          \\
  \end{tabular}
\end{table}

The Table~\ref{tab:fedComparison} synthesises the comparison. It is worth to say that the coupling that a style imposes varies inversely to its performances index (i.e., generated traffic and new member time), while it is proportional to the failure sensibility. 
The more coupled style, \textit{Publish~/~Subscribe with Replies}, is able to promote services with a minimal number of messages, ensuring that new members receive quickly previously promoted elements. The drawbacks are paid in terms of an elevate failure sensibility: if a single message is lost, there is a member that will not receive the entire promotion set.
On the other side there is the \textit{Publish~/~Subscribe} method, that requires a low coupling among different parties. Using the \textit{lease} concept, it is able to ensure a low failure sensibility, ensuring that the distributed coherence is hold. However, since it requires a periodic renew of each promoted element, this approach has a low efficiency.



\section{Performance Evaluation}
\label{sec:performance}
In order to prove the usability of our approach we evaluated its
performance in a realistic scenario, in which several enterprises
cooperate together by sharing services, subscribing to interests, and
organizing themselves in federations.

The overall delivery manager architecture requires many computational
resources, mainly due to the used registry and adopted application
server~\cite{ejb3spec}. This hampers the creation of a realistic
scenario, which demands the collaboration among many parties, each
one with its own delivery manager. In order to overcome unavailability
of such resources, we created a ``tiny'' version of the delivery
manager, able to simulate a realistic behavior without posing such
constraints on the available computational resources\footnote{The
``tiny'' version of the delivery manager doesn't interact  with a
registry (thus  all received elements are discarded) and adopts a
lightweight architecture, that doesn't guarantee an high level of
flexibility, extensibility and scalability.}. 
This way we are able to simulate several clients, able to generate a
significant amount of intra-registry interactions. 

In order to make those delivery managers cooperate together, we
created an simulation logic that automatically issues commands as a
human actor would do. The simulation logic periodically (i.e., every
5 minutes) performs a random action from the following set: 
\begin{itemize}
\item \textbf{marketplace}
  \begin{itemize}
  \item \textit{sharing} a service: a service is created and shared
    publicly. Each service has up to five specification facets, among
    which there is the WSDL one and optionally a QoS one. In the
    former is specified the method name, chosen among 100
    possibilities; in the latter we specify an hypothetical service's
    response time, varying between 0 and 100 tenth of
    seconds. Moreover, the simulation logic can also add other facets
    to simulate a comprehensive description of the service.
  \item \textit{subscription} to services: each simulation logic can
    perform a service subscription able to retrieve a shared
    service with a non-null probability. In particular, the generated
    interest leverages the WSDL facet to select a service with a
    particular method, chosen randomly within the 100
    possibilities. Moreover, it is possible to refine the scope by
    requiring that the matched service has a response time lower than
    a particular threshold. This way, each subscription has a match
    probability of circa 0.75 \%.
  \item \textit{sharing} an additional information facet: the
    simulation logic is able to generate and share additional
    information facets reporting a test suite of the service. This
    reports the completeness of the test (i.e., a number between 0 and
    1), measured in terms of functions covered.
  \item \textit{subscription} to additional information facets: once a
    service is received, it is possible to retrieve all its additional
    information facets containing a test suite having a completeness
    factor greater than a given threshold.
  \end{itemize}
\item \textbf{federation}
  \begin{itemize}
  \item \textit{join}/\textit{leave} of a federation: each delivery
    manager can join a federation, or leave one of the federations it
    belongs to. In order to enable an automatic behavior, we
    pre-created 100 federations with ps or psr communication style.
  \item \textit{promotion} of service / facet: a service or an
    additional information facet is created and shared in a federation
    the delivery manager belongs to.
  \end{itemize}
\end{itemize}

Obviously some sequences of actions are illegals (e.g. a delivery
manager can't promote a service in a not previously joined federation
or it can't join a federation it already belongs to). The simulation
logic is able to avoid such sequences, generating only valid sequences
of operation.

The goal of the first simulation we performed is to measure accurately
the matching time of the different interests, and foresee their
scalability. For this reason, we used seven computers in our
laboratory: the controlled environment guarantees that all of their
resources (including CPU, memory, and network) are dedicated to our
simulation, ensuring accurate results. 

\begin{figure}[ht]
  \centering
  \includegraphics[width=.6\columnwidth]{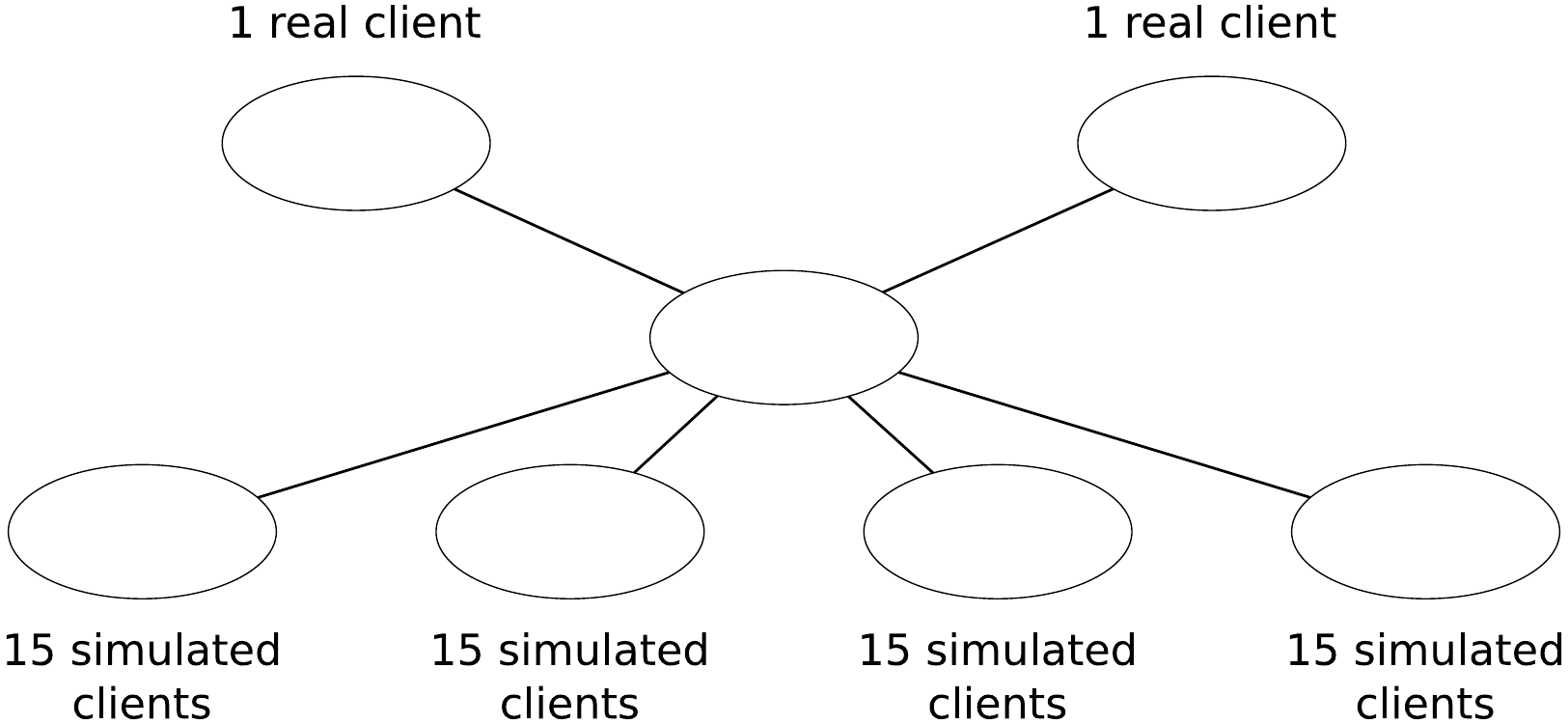}
  \caption{Topology of the controlled simulation.}
  \label{fig:localSim}
\end{figure}

Moreover, we designed a connection topology able to highlight the time
requested by the matching algorithm, that is exercised particularly on
brokers nodes: the resulting architecture is reported in
Figure~\ref{fig:localSim}.
The central machine hosts the broker, able to take trace of
subscriptions and forward messages on all the interested
subscribers. 
All of the other machines, act as traffic generator: two of them
have the full delivery manager version, while the remaining four
hosted 60 tiny delivery managers (15 instances per machine).
 
Instrumenting the broker, we measured the time required by a filter
to match a message; the average of those measurement are reported in
Table~\ref{tab:matching}.
\begin{table}[ht]
  \centering
  \caption{Matching time}
  \label{tab:matching}
  \begin{tabular}{r||rr|rr}
                  & \multicolumn{2}{c}{\textbf{positive match}} \vline & \multicolumn{2}{c}{\textbf{negative match}} \\
    \textbf{type} &  \textbf{time}    &   \textbf{throughput}   &   \textbf{time}     & \textbf{throughput}   \\ \hline
    Service       & 8,280.2 $\mu$s    &          0.44M msg/h    &   7,187.8 $\mu$s    &        0.50M msg/h    \\
    Facet AddInfo & 6,492.6 $\mu$s    &          0.55M msg/h    &      11.5 $\mu$s    &      313.04M msg/h    \\
    Federation    &     0.8 $\mu$s    &      4,500.00M msg/h    &       1.4 $\mu$s    &    2,571.43M msg/h     
  \end{tabular}
\end{table}

It is possible to notice that the average time requested to perform a
match is relatively small (below than 10 ms), but varies by some order
of magnitude depending on the complexity of the match.
Matching if a service complies to a user's requirements takes around 8
ms, mainly due the time requested by the standard XPath
checking. Anyway it is possible to notice a slightly decrease when the
outcome is negative: in this case the matching algorithm we
implemented uses a short-circuited boolean evaluation.

Verifying if a message containing a facet matches an end-user's
interest requires around 6.5 ms in the positive case, and 0.01 ms in
the negative case. This difference can be explained considering the
structure of the interest: it selects a set of facets (those compliant
to a given XPath) that refers to a particular service (specified using
its id). When a message arrives, the matching algorithm first checks
if the service id is the one desired, and then verifies if the XML
content is compliant to the given XPath. This allow us to prune a lot
of messages, not relevant for the end-user.
All messages directed to a particular federation, whose cooperation
style is based on REDS, uses a subject-based matching algorithm. It
performs a simple match on the federation identifier, and doesn't
require any check on the content of the message. For this reason the
matching time is very low, around 1 $\mu$s.

Our study continued analysing furthermore the matches required by the
marketplace, that involves an XPath constraint checking. 
For this reason, we studied the scalability of our proposal,
considering primarily how the system evolves when there are more
potential customers (i.e., the number of interest increases), and
after which are the consequences of having more service providers
(i.e., the number of messages increases).

\begin{figure}[ht]
  \centering
  \includegraphics[width=.6\columnwidth]{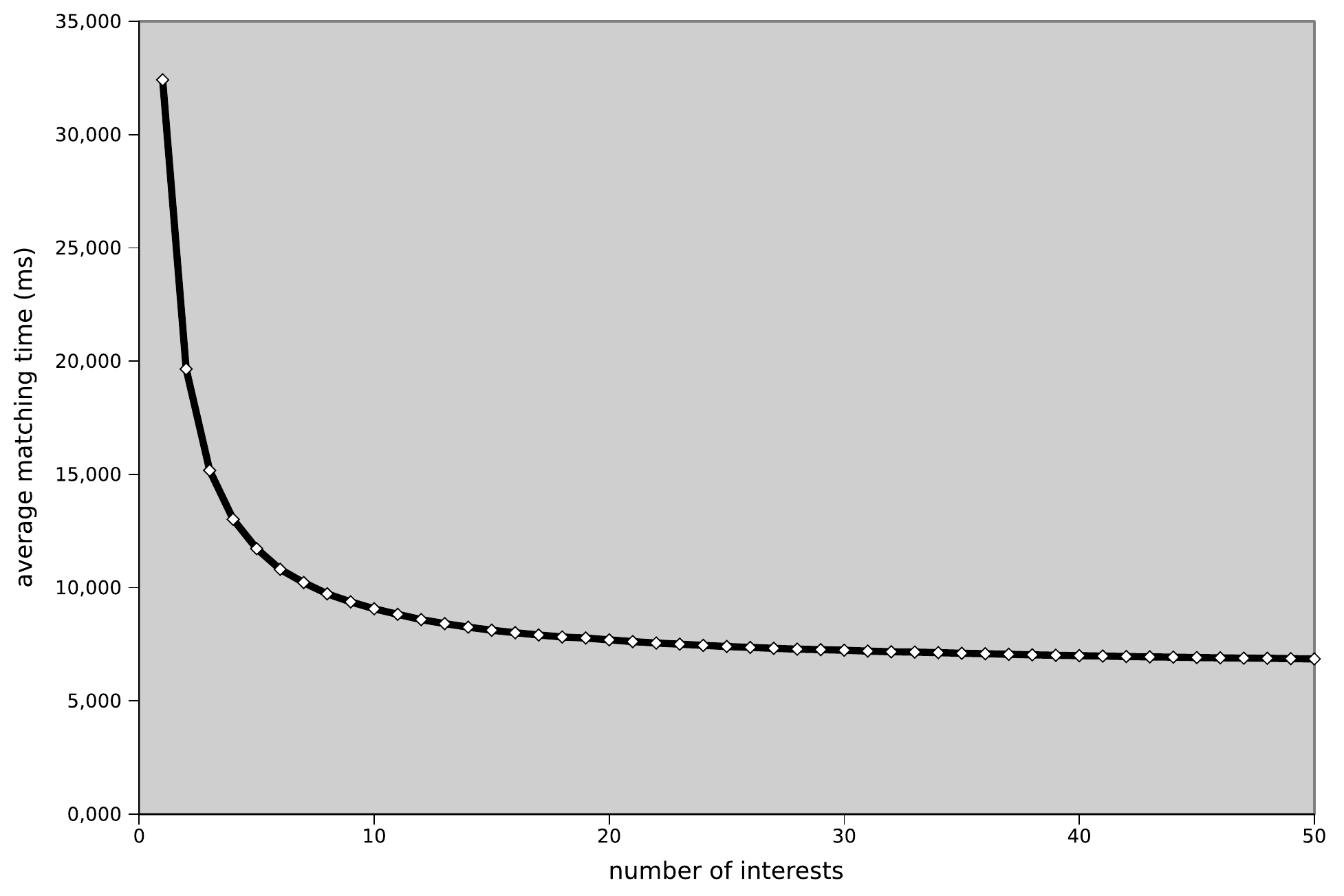}
  \caption{Marketplace interests: average matching time w.r.t. number of interest.}
  \label{fig:serviceInterestMatch}
\end{figure}

The chart present in Figure~\ref{fig:serviceInterestMatch} reports the
mean time that an interest requires to check whether a marketplace
message is interesting or not with respect to the number of
marketplace interests present in the broker.
With an high number of interests, the average matching time has an
asymptote value of 8.3 ms, that is the one reported in
Table~\ref{tab:matching}. It is also worth to notice that the first
time that an interest is used on a message, it is required more time
(slightly more than 30ms), due to some initial processing on the
message, such as the parsing of the XML content and the construction
of the relative Document Object Model (DOM). Subsequent matches on
that message can skip those phases, reusing the DOM created by the
first interest matched. 
Moreover it is possible to notice that the overall XML parsing and the
creation of its DOM graph is more complex than the XPath verification,
so our prototype can guarantee a good scalability with regard to the
number of subscriptions.

\begin{figure}[ht]
  \centering
  \includegraphics[width=.5\columnwidth]{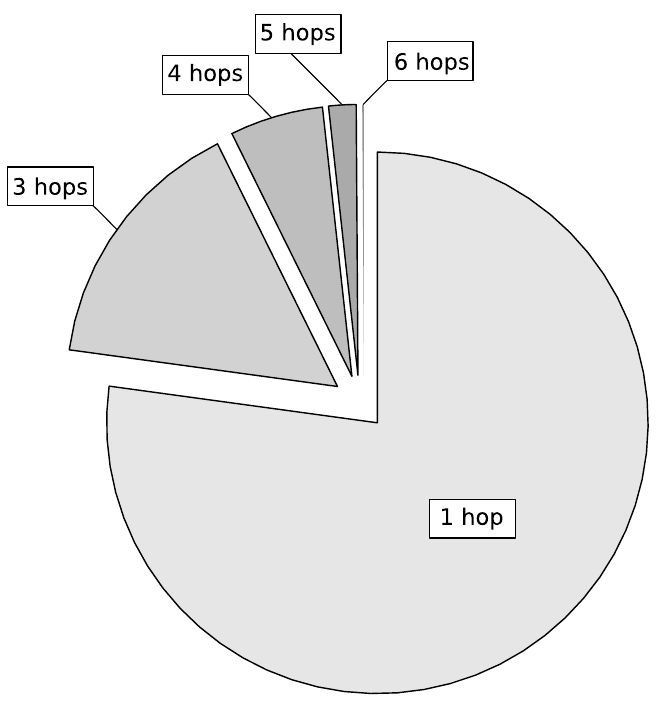}
  \caption{Number of hosts traversed by a message.}
  \label{fig:msgHops}
\end{figure}

In order to study the reaction of the system to an increase in the
number of service providers, we analyzed the effect of each single
promotions, in terms of number of brokers that a message pass
through. For this reason, for each messages published on the
publish~/~subscribe network\footnote{Within the distributed
  publish/subscribe systems, it is possible to have two main
  strategies: subscription forwarding and message forwarding. The
  former requires that all subscriptions are forwarded to each host,
  that checks whether a message matches to some interest. The opposite
  approach is to let each node to know only its own interests,
  broadcasting every message and the match it is performed only to
  delivery the message to the local clients. The former is useful
  whenever the size of exchanged messages are greater than the size of
  interests, while the latter is useful when interests are big and a
  client can perform frequently subscription and de-subscription. The
  delivery manager uses the first category of distributed
  publish/subscribe.}, we counted the number of hosts it passes
through, creating the chart in Figure~\ref{fig:msgHops}.
A significant amount of messages sent doesn't have any matching
interest, so them are discarded directly on the sender's node. This is
a peculiarity of the distributed publish/subscribe middleware that we
used, and allows us to guarantee that all traffic in the network is
relevant to at least one end-user. 
Considering the topology of the simulation, it is easy to understand
why a message doesn't traverse exactly two nodes: since there is a
central broker that connects all users, a message either can be
discarded, or it must traverse at least three nodes.
Moreover, it is possible to notice that the probability that $n$ hosts
receives a message decreases when $n$ augments; this means that
the algorithm used is able to detect some structure even if the
simulation we made is extremely random.
Combining these two consideration, it is possible to notice that our
proposal is able to minimize the effort required to manage the
messages created by a new service provider.

\begin{figure*}[ht]
  \centering
  \includegraphics[width=.9\textwidth]{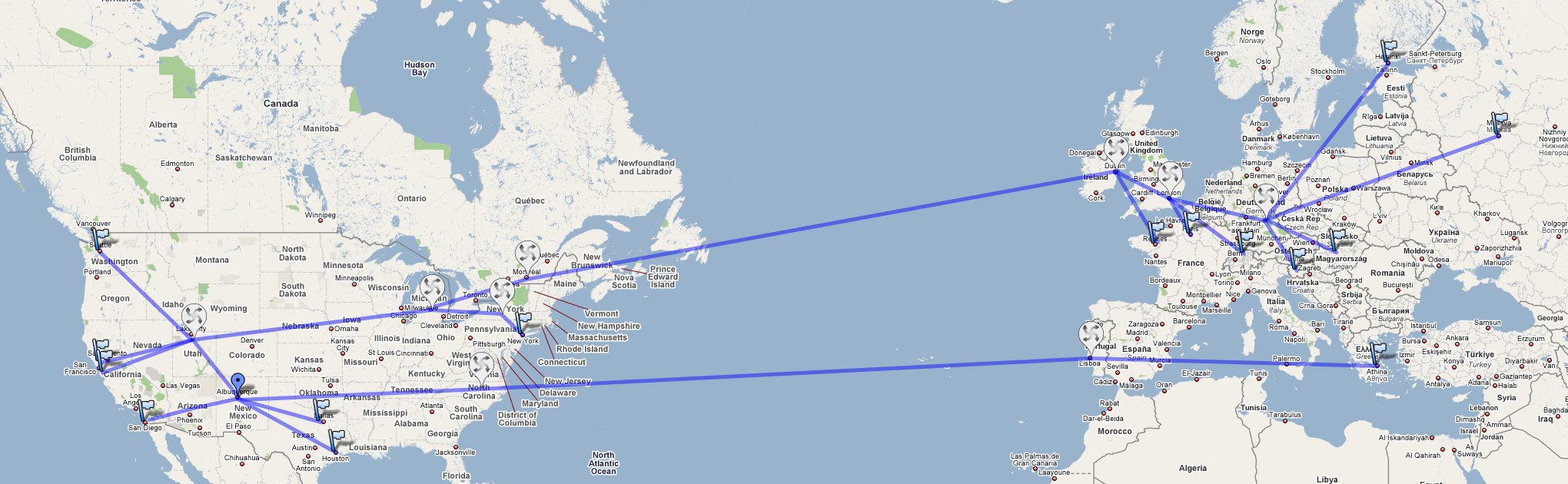}
  \caption{Topology of the planetLAB simulation.}
  \label{fig:topoPlanetlab}
\end{figure*}

The second set of simulations uses
\textit{PlanetLAB}~\cite{PlanetLAB}, giving us access to a set of
machines spread all over the world, with other potential users that
interacts with them. For this reason, the measured delays become
realistic, since are experienced variable delays due to the status of
the network, which cross-cut several countries. In this way we built
up a wide network of brokers, whose topology is shown in
Figure~\ref{fig:topoPlanetlab}; we connected our delivery manager (two
real and the other simulated) to the leaves of this network (in the
Figure, a flag indicates the presence of a delivery manager connected to that node). 
On top of this distributed network we run our simulation for 20 hours,
that involved 25 broker nodes, 2 real plus 21 simulated delivery
managers, 813 services, 3665 additional information facets, and 716,042
exchanged messages. 

\begin{figure}[ht]
  \centering
  \includegraphics[width=.6\columnwidth]{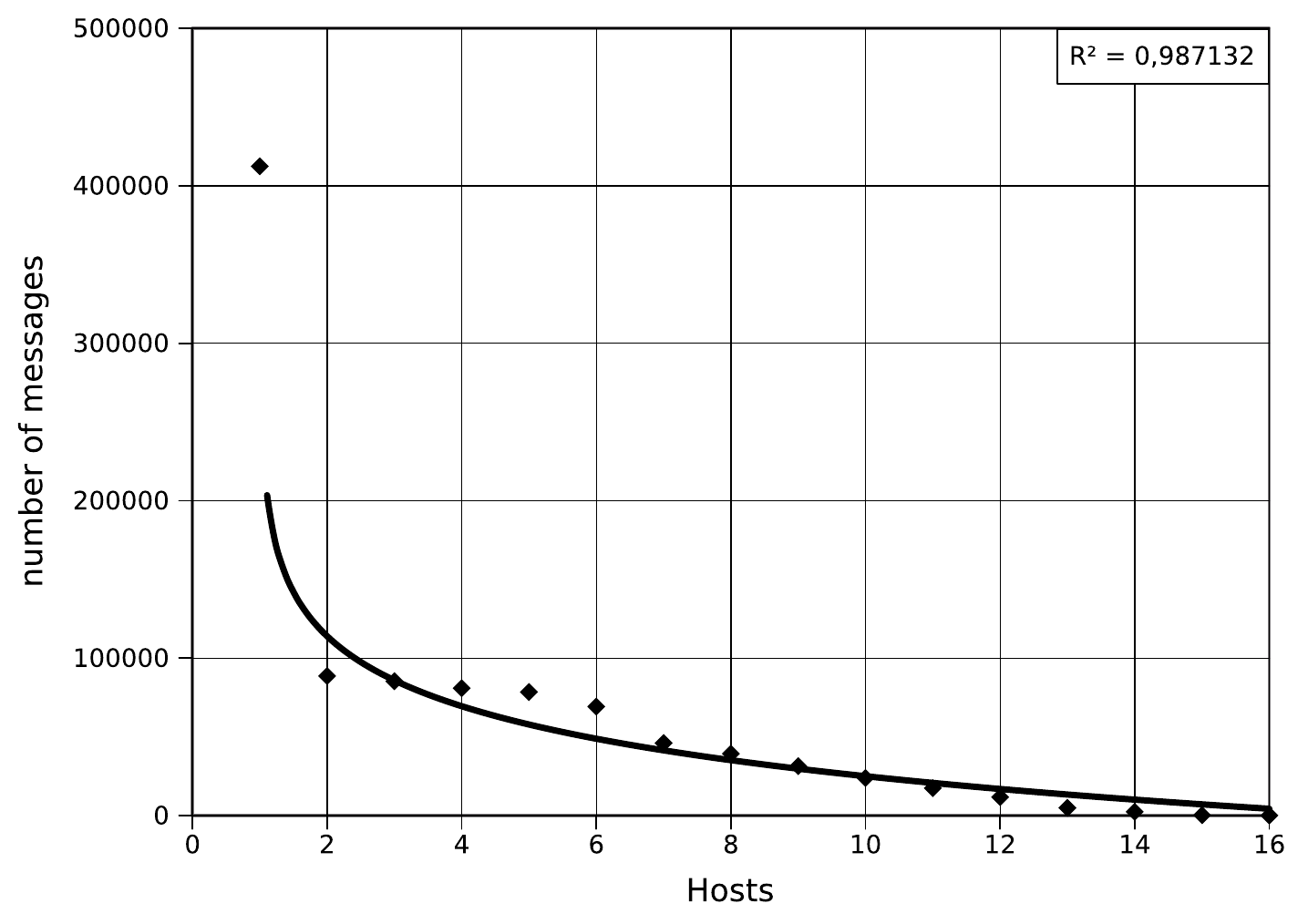}
  \caption{Number of hosts traversed by a message in PlanetLAB.}
  \label{fig:msgHopsPlanetLAB}
\end{figure}

Analysing the execution logs we firstly want to validate the
scalability with regards to the number of sent messages, leveraging on
the characteristics of a distributed publish/subscribe system such as
REDS. For this reason, for each sent message we counted the number of
traversed hosts, creating the chart reported in Figure~\ref{fig:msgHopsPlanetLAB}.
This chart strengthen our hypothesis on the scalability of the
proposal when new elements are shared. 
The first match is able to discard all of the uninteresting messages:
in our simulation tree fourth of the total exchanged messages are
pruned. It is worth to notice that the first match happens on
the sharing node, so everybody that wants to share something pays the
consequent set of comparisons.
The second consideration that it is possible to do is on the overall
shape of the chart: a negative logarithmic function fits well our
data, having an $ R^2 $ value close to 1. This fortify our conjecture,
ensuring that the probability that a new message pass through
\textit{n} hosts decreases with the augment of \textit{n}.

Finally, we want to measure the robustness, the efficiency, and the
performance of the overall delivery manager.
We reused the information extracted from the PlanetLab simulation to
evaluate both the marketplace, and the REDS-based federation
managers. 
Since the gossip-based federation manager requires at least one hundred of nodes
to work properly, we performed an ad-hoc simulation with 500 nodes for
its evaluation.

Regarding the marketplace, we measured a reliability of 67.21\%, and
we experienced an average delay of 43.84 s. It is worth to remember
that these values are obtained in a real environment: during the
simulation some links were broken and some node crashed, and we let
the overall system to self-adapt to it without providing any aid.

Regarding the federations, we tested each one of the three available
federation managers. The gossip-based one is the most reliable
federation manager, delivering successfully 99.44\%  messages; 
at the second place there is the one based on publish~/~subscribe,
with a delivery rate of 96.18\%, 
and finally there is publish~/~subscribe with replies, that delivered
successfully 76.49\% messages.

Afterwards we measured the number of exchanged messages compared with
the number of promotion made; these values are reported in
Table~\ref{tab:fedMessages}. It is possible to notice that the most
efficient manager is the PSR one, while gossip requires several
messages\footnote{REDS-based federation manager can leverage on the
  multi-cast capability of the middleware, being able to send one
  message addressed to multiple recipients. However, the Gossip-based
  federation manager uses the raw TCP socket, and thus has to send a
  message for each recipient.}.

\begin{table}[ht]
  \centering
  \caption{Number of messages sent for each federation type.}
  \label{tab:fedMessages}
  \begin{tabular}{c|rrr}
    type   & \# messages & \# events & msg / event \\ \hline
    psr	   &     582    &   291          &     2.00 \\
    ps	   & 124,908    &   298	         &   419.15 \\
    gossip & 212,398    &   150          & 1,415.99 \\
  \end{tabular}
\end{table}

The last result we measured is the average time to transfer an
element in a federation, that is 55.255 s for publish~/~subscribe, and
9.165 s for publish~/~subscribe with replies. As expected,
PSR is faster than PS, requiring a transfer time that is 16,59\% than
the PS one.


\section{Conclusions}
\label{sec:conc}

This article presents \textit{DIRE}, an approach for the cooperation
and federation of distributed heterogeneous registries based on the
publish and subscribe paradigm. The single ``logically'' centralized
\emph{dispatcher} acts as common reference for the registries that
want to communicate, but it maintains a high degree of independence
among the registries.  Each entity is free to decide what information
---and thus what services--- it wants to share within the community by
\textit{publishing} it through the dispatcher. Similarly, they can
also decide the services they are interested in by
\textit{subscribing} to particular service types. Federations can be
set among registries to support the broadcast of information among the
elements that belong to the federation.  The whole approach is also
based on a dedicated service model to provide powerful and flexible
descriptions of services and to support the creation of powerful
filters for sophisticated subscriptions. The proposed model is
independent of the technology of the registries that form the
community. 



\appendix
\section{Publish~/~Subscribe}
\label{sec:pubSub}
In this Section we will give an overview of publish/subscribe systems and we will illustrate the main features of REDS, the framework adopted in our work to support the publish/subscribe communication style.

\subsection{Publish/subscribe systems}
\label{sec:pubsubsystems}
Coordination-based systems \cite{tanenbaum2002dsp} aim to coordinate the activities of their components, inherently distributed. The most interesting aspect of this approach is that it doesn't take into account the single computation of each process, but it handles all the communication and cooperation between processes. In this kind of systems, the focus is on how coordination between processes takes place.

Meeting-based systems \cite{tanenbaum2002dsp} specialize coordination-based systems, including a meeting concept in which processes temporarily group together to coordinate their activities. Publish/subscribe systems are, in their turn, a specialization of meeting-based systems, implemented by means of subject based messaging. These applications are widely used to easily disseminate information to multiple users who may be interested in some or all of the information available. Moreover, they also give users the possibility to modify received data and subsequently advertise them.

Publish/subscribe applications \cite{PubSubWithRep} \cite{DistributedPS} are organized as a set of distributed components, \textit{publisher}s, \textit{subscriber}s and brokers. Publishers advertise information by publishing messages, which could represent events such as, news, available services, etc. Subscribers declare their interest for some kinds of information doing subscriptions. The broker provide access points to clients to advertise information and to subscribe for notification of interest. It is in charge of collecting subscriptions and routing messages from publishers to the interested subscribers.

Subscriptions can be of two types \cite{libroIBM}: 
\begin{itemize}
\item \textit{Content based} subscriptions express constraints on the message content, through patterns or filters. For example, a subscriber can be interested in services that have a cost less than 10\$. This selection process also can be used to optimize communication within the network. More specifically the broker may be asked to apply a filter to the contents of published messages, such that it will deliver only those that contain specified data values. The selection process may also be asked to look for patterns of multiple messages, such that it will deliver only sets of messages associated with that pattern of event occurrences (where each individual event occurrence is matched by a filter). 
\item \textit{Topic-based} subscriptions express the interest on messages belonging to a specific subject, that is a meta-data associated to messages. For example, a subscriber could be interested in all services involved in the area of clothing.
\end{itemize}

Brokers should provide two functionality \cite{DistributedPS}: 
\begin{enumerate}
\item \textit{Message selection}. Brokers carry out a selection process to determine which of the published messages are of interest to which of its subscribers.
\item \textit{Message delivery}. Brokers correctly route the matching messages to all the interested subscribers.  
\end{enumerate}
A typical trade-off \cite{DistributedPS} of publish/subscribe systems is between the complexity of the matching function executing during selection and the scalability of the routing. The complexity of the matching function depends on the language used to advertise messages and do subscriptions. In fact as soon as the expressive power of the language increases, the complexity of the processing to match subscriptions raises too.

The interaction model \cite{PubSubWithRep} of publish subscribe systems is implicitly \textit{anonymous}, since it provides loosely reference between publisher and subscribers. In these applications publishers haven't any knowledge about the identity of subscribers and, in the same way, subscribers don't need to know anything about the publisher. Moreover subscribers are in their turn independent among each others. 
Publish/subscribe communication is \textit{asynchronous}, because publishers and subscribers operate in parallel without synchronizing during communication, and \textit{multi-point}, since each message can be delivered to many interested subscribers. The interaction protocol is \textit{stateless}, since only subscriptions are persisted in the brokers while messages are sent only to those components that have subscribed before the messages are published.
The way in which communication takes place depends on the subscriptions content and where they were submitted. To ensure location transparency, multiple brokers can  be connected together to exchange published messages. In this way, subscribers don't need to do subscription in the same broker in which message was published. 
 
One of the main advantage of this kind of systems is the strong decoupling between publishers and subscribers, which greatly reduces the effort required to modify the application architecture at run-time by adding or removing components. But one of the main disadvantages of centralised publish/subscribe systems is their low scalability; if it is required to cope with large numbers of components, it is possible to adopt distributed solutions.

\subsection{REDS - A Reconfigurable Dispatching System}
\label{sec:REDS}
REDS - (REconfigurable Dispatching System) \cite{reds} is a framework to build publish/subscribe applications for large, dynamic networks. REDS provides an infrastructure of components, with clearly defined interfaces, to build a distributed dispatcher organized as a set of brokers linked in an overlay  network, which collaborate to route messages from publishers to subscribers.

Each REDS broker provides the message selection and message delivery functionality through the overlay and the routing layer.
The former enables mechanisms to maintain broker overlay network when the underlying topology changes. This mechanism guarantees that broker can exchange message while network connection changes.
Routing layer encapsulates routing functionality: maintains subscriptions in memory, perform matching function to detect matching subscriptions when a message arrives and select the best strategy to route to message to interested subscriber. It also guarantees that routing information, and, in particular, the content of subscription tables, remains consistent when the topology of the dispatching network changes. 

With respect to other publish-subscribe middleware REDS provides several innovations:
\begin{itemize}
\item \textit{Content-based routing} \cite{migliavacca2007aps}. Routing can be done by propagating information about clients' subscriptions along the same dispatching tree and subsequently using such information to route messages only toward interested brokers, i.e., those that have at least one subscribed client attached to them, along a single acyclic overlay.

\item \textit{Replies} \cite{PubSubWithRep}. REDS provides bidirectional communication allowing brokers to reply to particular messages tagged as \texttt{Repliable}. Every REDS brokers keep track of the transit of those messages and store routes followed back by replies. Implementing replies at middleware level, REDS brokers are able to track the number of expected replies for each message and check if and when all of them have been received.

\item \textit{Self-healing} \cite{frey:phd}. REDS allows nodes of the network to self-repair to react to  run-time reconfiguration of the dispatching network, either to react to changes in the underlying physical network or to adapt it to the application's needs, e.g., to balance the traffic load, or to change the number of brokers and their connectivity. The adopted repair strategies minimize the changes that impact on the content based routing.

\item \textit{Self-organization} \cite{migliavacca2007aps}. Since the cost of routing depends on the distance between subscribers and the broker where the interesting message was published, it can increase when the message must traverse a large number of brokers to reach its recipients. To overcome this limitation the dispatching network is able to periodically reconfigure its topology, depending on the subscriptions, at runtime, to reduce the overall routing cost. 
\end{itemize}




\bibliographystyle{IEEEtran}
\bibliography{biblio}




\end{document}